\documentclass[manuscript]{aastex62}
\usepackage{graphicx}
\usepackage{epstopdf}

\shorttitle{AstroSat UVIT Detections of Chandra X-ray Sources in M31}
\shortauthors{Leahy \& Chen}

\begin{document}

\title{AstroSat UVIT Detections of Chandra X-ray Sources in M31}

\author{D. A. Leahy and Y. Chen}
\affil{Dept. of Physics $\&$ Astronomy, University of Calgary,
Calgary, Alberta, Canada T2N 1N4}

\begin{abstract}
An ultraviolet survey of M31 has been carried out during 2017-19 with the UVIT instrument onboard the AstroSat Observatory.
Here we match the M31 UVIT  source catalog \citep{2020ApJS..247...47L}  with the Chandra source catalog \citep{2016MNRAS.461.3443V}. 
We find 67 UVIT/Chandra sources detected in a varying number of UV and X-ray bands.
The UV and X-ray photometry is analyzed using powerlaw and blackbody models.
The X-ray types include 15 LMXBs and 5 AGNs. 
Cross-matches with catalogs of stars, clusters and other source types yield the following. 
20 of the UVIT/Chandra sources match with M31 globular clusters and 9 with foreground stars.
3 more globular clusters and 2 more foreground stars are consistent with the UVIT source positions although outside the Chandra match radius of $1^{\prime\prime}$. 
The UV emission of the UVIT/Chandra sources associated with globular clusters is consistent with emission from blue horizontal branch stars rather than from the X-ray source.
The LMXBs in globular clusters are among the most luminous globular clusters in M31.
Comparison with stellar evolutionary tracks shows that the UVIT/Chandra sources with high UV blackbody temperatures are consistent with
massive (10 to 30 M$_{\odot}$) stars in M31.
\end{abstract}


\section{Introduction}\label{sec:intro}

M31 is our closest neighboring large galaxy and is a spiral similar in many ways to the Milky Way.
Studies of large numbers of stars have been done in detail for our own Galaxy, but have uncertainties
related often to uncertain distances or to strong extinction.
The advantage of studying objects in M31 is that it is at a well known distance (783 kpc, \citealt{2005McConnachie}).

A  UV instrument was launched on 28 September 2015 as part of the AstroSat mission \citep{singh+2014}.
AstroSat  has four instruments, covering near and far ultraviolet (NUV and FUV) 
with the UVIT telescope, and soft through hard X-rays with the SXT, LAXPC and CZTI instruments.
The UVIT telescope and its calibration are described in \citet{2020AJ....159..158T}, \citet{2017Tandon2}, 
 \citet{2017Tandon1}, \citet{2011Postma} and references therein.
The UVIT observations have high spatial resolution ($\simeq$1 arcsec), 
thus have the capability of resolving individual stellar clusters and a large number of individual stars in M31.

This paper presents an analysis of sources in M31 detected in both UV and X-rays. It  utilizes the M31 
UVIT catalog \citep{2020ApJS..247...47L} and M31 Chandra catalog \citep{2016MNRAS.461.3443V}. 
We briefly describe the observations in section \ref{sec:obs}.  
The source cross-identification and the analysis of the photometry is presented 
in section \ref {sec:analysis}. 
We discuss source properties in section \ref{sec:results} and 
summarize the results in section \ref{sec:summary}.

\section{Observations}  \label{sec:obs}

UVIT is comprised of two 38 cm telescopes, one for far ultraviolet (FUV) (130 to 180 nm) wavelengths and 
one for near ultraviolet (NUV) (200 to 300 nm) and visible (VIS) (320 to 550 nm) wavelengths.
The FUV, NUV and VIS channels each have a number of filters with different bandpasses.
The pixel scale for UVIT images is 0.4168 arcsec per pixel. 
Point sources in the UVIT images have FWHM $\simeq$1 arcsec in the FUV and NUV channels. 
To create the M31 UVIT catalog  a new analysis of the 
UVIT photometric and astrometric errors  was carried out \citep{2020ApJS..247...47L}.
The following filters were used for the M31 UVIT survey: CaF2 (123 to 173 nm), BaF2 (135 to 173 nm), 
Sapphire (146 to 175 nm), Silica (165 to 178 nm), NUVB15 (206 to 233 nm) and NUVN2 (275 to 284 nm). 

The UVIT instrument \citep{2017Tandon2} consists of an FUV telecsope and a NUV/VIS telescope.
The latter has a beam splitter to direct the NUV and VIS light onto 2 different detectors. 
The VIS channel is used for spacecraft pointing, so normally science observations are carried out simultaneously in FUV and NUV channels. 
The NUV channel failed in early 2018, so that science observations since then have been carried out in FUV only.
Whenever two or more FUV filters or two or more NUV filters are utilized, the observations must take place
at different times. 
The BJD (Barycentric Julian Dates) of the start times of the observations of M31 are given in Table 1
of \citet{2020ApJS..247...47L}.

The field of view of each UVIT telescope is $\sim$28 arcmin in diameter, thus to cover the large area of M31 
19 different fields (listed in Table 1 of \citealt{2020ApJS..247...47L}) were required. 
For M31, the observation times are clustered  in 3 $\sim$20-60 day intervals around BJD2457700, BJD2458080 and BJD2458400, separated by $\sim$1 year.
The observations in different FUV and NUV filters for a given field were observed within 0.2 to 0.5 days,
with exposure times of between 1900 and 15000 s. 
There were no repeated observations of any field in any single filter.

The FUVCaF2 filter was used for 18 of the 19 fields of M31; NUVN2 filter for 9 fields; NUVB15 filter 
for 10 fields; FUVSilica filter for 10 fields; FUV Sapphire filter for 6 fields; and FUV BaF2 filter for 1 field. 
This, combined with the wide bandwidth of FUV CaF2 filter (500 nm) compared to the other filters 
meant that the CaF2 observations detected the largest number of sources.
We used a threshold of 3 sigma excess above local background to identify potential point sources. 
Then we perform a PSF (point spread function) fit to the source and kept the source  
if total counts in the fit was more than 8 sigma above background.
Thus the likelihood of false detections is small among the $\sim$30,000 sources detected
in the CaF2 filter, or among the total number of detections of $\sim$70,000 for all
filters.

The position uncertainties for UVIT were determined as follows.
The images were registered to optical position calibrators from Gaia. 
The standard deviation of the resulting Gaia-UVIT offsets ranged from 0.15 arcsec to 0.7 arcsec 
(Table 2 of  \citealt{2020ApJS..247...47L}).
Then the offsets between sources in the same field measured in different filters were 
measured (Fig. 15 of \citealt{2020ApJS..247...47L}), showing a peak for real matches at separation of 0 to 0.4 arcsec, and a rise for separations $>$2 arcsec consistent with
the expected number of accidental matches.

\begin{figure}
\plotone{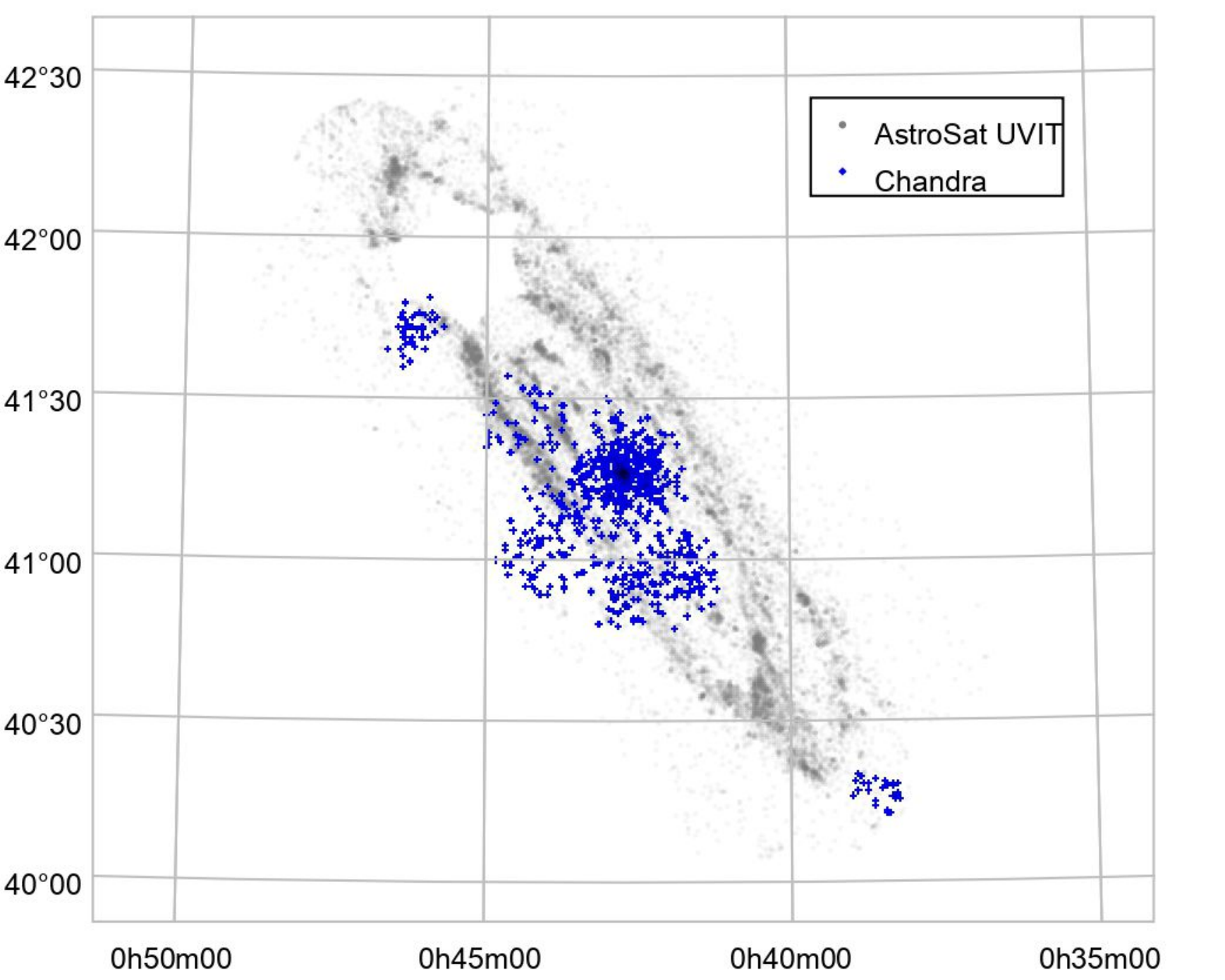}
\caption{Plot of the positions all M31 UVIT sources in the CaF2 band (grey dots, from the catalog of \citealt{2020ApJS..247...47L}) and of the Chandra sources (blue crosses, 
from the catalog of \citealt{2016MNRAS.461.3443V}). 
}
\label{fig:SourceMap}
\end{figure}

A description of the Chandra source catalog for M31 is given in \citet{2016MNRAS.461.3443V}.
The Chandra images were registered to the CHFT i-band image of M31, which was corrected 
to match the Two Mirocn All Sky Survey, which is accurate to better than 0.2 arcsec.
The position errors listed in the source catalog were taken as standard deviations of
the PSF in the extraction region and are typically 0.3 arcsec.
A map of the positions of the 31337 sources detected in the UVIT CaF2 filter is shown in Figure~\ref{fig:SourceMap}.
Overlaid on the UVIT source positions are the positions of the 935 X-ray sources in the Chandra catalog.

\section{Analysis}\label{sec:analysis}

\subsection{Cross-matching sources}

In order to cross-match the catalogs, we  compared the position uncertainties of Chandra and UVIT sources and determined an appropriate match radius. 
In the Chandra catalog, position uncertainties are typically $\sim0.3''$, with more than $80\%$ below $0.7''$. 
The UVIT position uncertainties close to the bulge (where most Chandra sources are located) are on average $\sim0.2$-$0.5''$. 
Therefore, we expect the majority of Chandra sources and UVIT sources that are correlated to be separated by $\lesssim 1.0''$ of each other. 
Some UVIT and Chandra sources have position uncertainty above $1''$, meaning that a separation of $\sim2''$ is also possible. 

\begin{figure}
\plotone{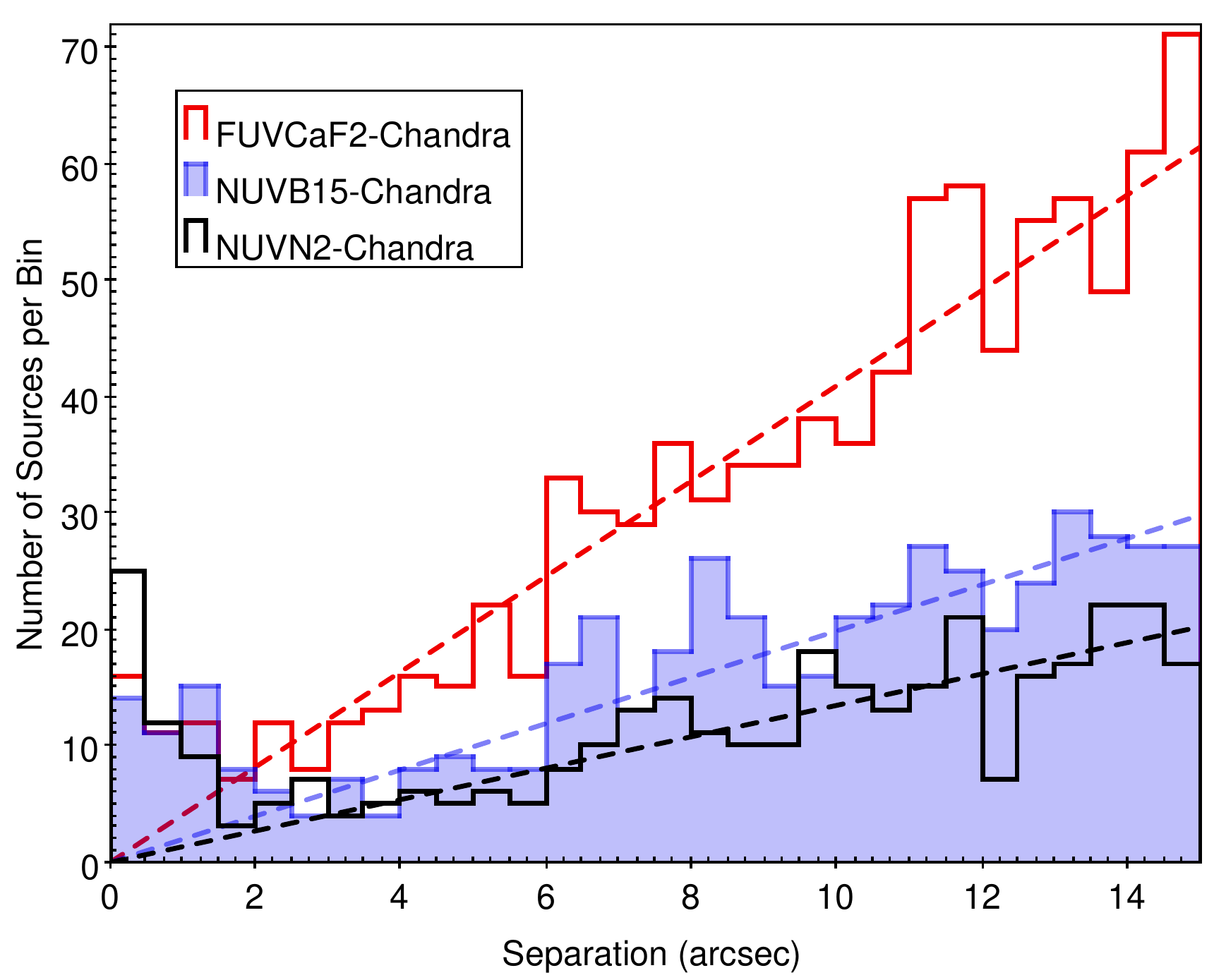}
\caption{Distribution of separations between UVIT sources and Chandra sources. 
Linear fits (dashed lines) to the distributions for separations $>3.0''$ are plotted as dashed lines, and extrapolated to smaller separations.}
\label{fig:SepDistri}
\end{figure}

In Figure~\ref{fig:SepDistri}, we plot the separation distribution between FUVCaF2 sources and Chandra sources; 
between NUVB15 sources and Chandra sources; and between NUVN2 sources and Chandra sources. 
We exclude FUVSapphire and FUVSilica sources from the analysis of separation distributions because there were not enough of them. 
All three distributions become approximately linear above a separation of $3.0''$, where UVIT and Chandra sources are no longer correlated. 
A linear function for number of matches per unit radius, $r$, is expected for a random distribution of
sources because the differential increase in area is $2\pi rdr$. 
We fit a linear function to each distribution using only points above a separation of $3.0''$,
 plotted as dashed lines in Figure \ref{fig:SepDistri}. 
We then extrapolated the line of best-fit to small separations to estimate the number of accidental matches. 

According to the line of best-fit, accidental matches are negligible for separations between $0$-$1.0''$ but become significant for separations around $2.0$-$2.5''$ for all three distributions.  
We assume that FUVSapphire and FUVSilica sources follow the same pattern as well. 
Therefore, to minimize accidental matches while keeping as many real matches as possible, 
we chose $2.0''$ as the match radius for cross-matching UVIT and Chandra sources, and took only the closest match within the radius. 
In total, we matched 67 UVIT sources with 67 Chandra sources. 1 source is detected in 5 UVIT bands;
17 sources in 4 bands; 24 sources in 3 bands; 8 sources in 2 bands; and 17 sources in 1 band. 
The 67 UVIT/Chandra sources are listed in Table \ref{tab:MatchIndices}.

From Figure~\ref{fig:SepDistri} we estimate the number of accidental/spurious matches between the different UVIT filter bands with separation $<2^{\prime\prime}$:
for CaF2  $\sim$9; for NUVB15 $\sim$5; and for NUVN2 $\sim$3. 
We separated the total matches into those with separation $<1^{\prime\prime}$ and those with 1-2$^{\prime\prime}$.
The numbers are: for CaF2 27 in $<1^{\prime\prime}$ plus 15 in 1-2$^{\prime\prime}$;
for NUVB15 25 in $<1^{\prime\prime}$ plus 19 in 1-2$^{\prime\prime}$;
for NUVN2: 37 in $<1^{\prime\prime}$ plus 9 in 1-2$^{\prime\prime}$.
To indicate the higher probability of accidental match,
sources that match within 1-2$^{\prime\prime}$ are marked in Table 1 with an asterisk for the appropriate UVIT band.
The fraction of total matches wth separations 1-2$^{\prime\prime}$ is highest (43\%) for NUVB15 and lowest (20\%) for NUVN2. 
The expected fraction of accidental matches with separations 1-2$^{\prime\prime}$ is highest for CaF2
(6 of 15) and smaller for NUVB15 (3.5 of 19) and NUVN2 (2 of 9).

\subsection{UV spectrum analysis}

The color-color diagram using FUVCaF2, NUVB15, and NUVN2 magnitudes is shown in Figure \ref{fig:colorcolordiag1}. 
41 of the 67 matched sources were detected in these three UV bands. 
The extinction curve has a peak at 4.59 $\mu^{-1}$ (218 nm) with a half-width of $\sim20$ nm \citep{2007ApJ...663..320F}. 
This peak coincides with the NUVB15 filter. As a result, the FUVCaF2-NUVB15 color 
becomes bluer and the NUVB15-NUVN2 color becomes redder as extinction increases, as seen in Figure 3.

A series of black bodies of different temperatures, and a series of stellar models (from \citealt{2004Castelli}) 
that correspond to different main-sequence stars, are plotted in Figure~\ref{fig:colorcolordiag1}. 
Three different extinctions\footnote{We use the extinction curve of \citet{2007ApJ...663..320F}.} 
$A_V = 0$, $0.21$ and $1.21$ are connected by dotted lines.
$A_V = 0.21$ is the expected foreground extinction to M31 from our Galaxy.  
The data points right of the model points could be caused by higher extinction than $A_V = 1.21$.

The CK models for stars are 
redder than blackbodies in NUVB15-NUVN2 color for some T$_{eff}$ and bluer for other T$_{eff}$. 
A bigger difference is seen in  FUVCaF2-NUVB15 color:
The CK models are much redder for T$_{eff}\lesssim 8000$ K and not much different in color
for higher T$_{eff}$.
It is seen that the data 
are roughly consistent with blackbodies with temperatures in the range $\sim5700$ K to 30000 K, 
or with stellar spectra  with temperatures in the range $\sim7200$ K to 15000 K.
The reason is that the UV emission of stellar spectra is more sensitive to temperature than for blackbodies.
For bluer sources (FUVCaF2-NUVB15$<$2) the blackbodies and stellar models cover nearly the same region. 

Figure \ref{fig:colorcolordiag2} is an expanded plot, where we show the 28 sources at the bottom left of Figure \ref{fig:colorcolordiag1}.  
On this diagram, we mark a series of points using template galaxy spectra as well as power law 
continuum of AGNs and non-AGN Lyman-break galaxies\footnote{Galaxy templates are from the Kinney-
Calzetti Spectal Atlas of Galaxies at STScI; AGN and Lyman-break galaxy templates are from 
\citet{2011ApJ...733...31H}.}. 
The lines joining black bodies with $A_V = 0.21$  (blue dashed line) and $A_V = 1.21$  
(cyan dashed line) are shown\footnote{
$A_V = 0.21$ is the Milky Way foreground reddening and we estimate internal extinction in M31 of $\sim$0 to $\sim$1.
The galaxy templates include internal reddening, depending on galaxy type. Thus all three contributions to reddening are included.
}.
This illustrates that the region covered by blackbodies (from Figure 
\ref{fig:colorcolordiag1}) overlaps that of galaxies,
which makes it hard to distinguish stars from galaxies in the UV color-color diagram.
Only bulges and Sc galaxies are significantly offset from the stars region: they are 
redder in NUVB15-NUVN2 color, yet similar in FUVCaF2-NUVB15 color.

For the 41 matched sources with three or more UV magnitudes,
we fit their UV magnitudes using: i) a black body (BB) function with parameters 
temperature, radius, and $A_V$; and ii) with a power-law (PL) function with parameters 
normalization,  PL  index, and $A_V$.
For the PL, we use the form
$F_{\nu} = norm\times (\nu/\nu_0)^{-\alpha}$, with $\nu_0=1.5\times10^{15}$ Hz
and $\alpha$ is the PL energy index. 
We use least-squares fitting, minimizing $\chi^2=\sum_{i=0}^{n}(d_i-M(p_j))^2/(e_i^2)$, with $d_i$ 
and $e_i$ data and error for filter i and $M(p_j)$ the model with parameter set $p_j$.   

Fig.~\ref{fig:spectra} shows spectra for four of the sources and their best-fit BB and PL models. 
The source in the upper-right panel has a power law index of 8.1, which is unphysical. The probable correct
interpretation is a thermal model, where the steep slope is caused by fitting a power law to the Wien tail of a thermal spectrum.
The best-fit parameters and $\chi^2$ values for each source, for both BB and PL fits,  are 
shown in Table~\ref{tab:MatchBBfit}.
The $90\%$ errors in parameters were calculated by varying the fit parameters to obtain $\chi^2$
equal to the best-fit value plus 6.21 for the case of 3 parameters (see Chapter 15 in \citealt{2002nrca.book.....P}).

For the 8 sources with only two magnitudes, we assume a fixed extinction of $A_V=0.30$
and fit the 2 data points with two parameter BB or PL models to obtain the parameters
(temperature, radius for BB or normalization and index for PL).
The $90\%$ errors were obtained as above except that for the case of 2 parameters
we use best-fit value plus 4.61. 
Sources with only one magnitude have no fit (no index or $\chi^2$) and are not listed in 
Table~\ref{tab:MatchBBfit}. 

\subsection{UV and X-ray comparison}

Each source in the Chandra catalog has a summed soft band ($0.5$-$2.0$ keV) flux and a summed hard band ($2.0$-$8.0$ keV) flux. 
We use the ratio of the two fluxes to obtain an X-ray PL energy index, $\alpha_{X}$.
Results are shown in the tenth column of Table~\ref{tab:MatchIndices}. 
Some UVIT-Chandra sources have no measured hard band flux (given as $-9.99$ in the Chandra catalog). 
For these sources, we were unable to obtain a powerlaw index, so those entries are left blank in the table.

For each matched source, we utilise the UV flux and the X-ray flux to obtain an UV to X-ray PL 
energy index, $\alpha_{UVX}$. 
We calculate the index by taking the ratio between the summed soft band X-ray flux and the UV flux in the band with the highest energy. 
The index for each source is given in column eleven of Table~\ref{tab:MatchIndices}.
For both X-ray PL index and UV to X-ray PL index, the error is obtained from the photometric errors using error propagation.

\section{Results}\label{sec:results}

\subsection{Identifications of the UVIT/Chandra sources}

Table \ref{tab:MatchIndices} lists previous identifications of source types in X-ray and optical.
In X-ray, \citet{2016MNRAS.461.3443V} identified a subset of their sources as LMXBs or as
AGNs. 
The LMXBs were identified using \citet{2011A&A...534A..55S}.
The AGNs were identified by matching to the background galaxies and AGN from 
\citet{2015ApJ...802..127J}, quasars from the LAMOST survey 
\citep{2015RAA....15.1438H} 
and AGN listed in NED, SIMBAD and SDSS DR12.
15 of our 67 sources have LMXB identifications  and 5 have AGN identifications.
\citet{2018Sasaki} gave X-ray identifications for the XMM-Newton survery of the northern disk of 
M31\footnote{Matching to the catalog of \citet{2018Sasaki} used a $3^{\prime\prime}$
 cutoff (2$\sigma$ position error).}.
Their X-ray sources were classified as: foreground star (fgstar), X-ray binary (XRB), 
supernova remnant (SNR), Galaxy, or hard, using a combination of methods. They indicated 
candidate sources by putting the classification in angle brackets (e.g. $<$XRB$>$).
2 of our sources are XRB, 4 are $<$XRB$>$, 2 are fgstar, and 1 is $<$SNR$>$.

In optical, \citet{2015ApJ...802..127J} identified stellar clusters and background galaxies
in the PHAT survey, with 2753 clusters and 2270 galaxies identified\footnote{Matching 
to the catalog of \citet{2015ApJ...802..127J} used a $1.5^{\prime\prime}$
 cutoff (2$\sigma$ position error).}.
In Table \ref{tab:MatchIndices} we mark the 13 UV/X-ray sources 
identified as stellar clusters by C in the last column. 
None of our UV/X-ray sources matched background galaxies from \citet{2015ApJ...802..127J}.
\citet{2018ApJS..239...13W} compared Chandra and HST observations of M31 to study
X-ray sources\footnote{Matching to the catalog of \citet{2018ApJS..239...13W} used a 
$2.0^{\prime\prime}$ cutoff (2$\sigma$ position error).}. 
They identified the X-ray sources as point source (labelled p),
background galaxy, star cluster (c), foreground star (f),  SNR (s), or no PHAT counterpart (n).
We include their classification in the last column of Table \ref{tab:MatchIndices}. 
Our list of UV/X-ray sources includes 2 star clusters, and 1 each of p, f, s and n categories.
We also matched to their list of HMXB candidates and found none of our sources matched those.

The sources in Table 1 were searched for identifications using the VizieR catalog search tool (http://webviz.u-strasbg.fr/viz-bin/VizieR)
with a search radius of $2^{\prime\prime}$ of the Chandra position. 
Associations with separation $1-2^{\prime\prime}$ are marked with an asterisk in Table 1. 
Those with separation $1-2^{\prime\prime}$ are probably not associated with the X-ray source, and some with separation $<1^{\prime\prime}$ are chance alignments. 
Those were kept because of the larger UVIT source position errors: some could be associated with the UVIT sources although most are likely chance alignments.
Those found in the Gaia DR2 catalog \citep{2018AA...616A...1G} with Gaia distances \citep{2018AJ....156...58B} are indicated in Table 1 by GD or GD*. 
If the UVIT photometry was consistent with being a foreground star, the source was marked as GDfg or GDfg*. 
Sources identified as globular clusters in M31 (\citealt{2016ApJ...824...42C}, \citealt{2016ApJ...829..116S}) are marked as GC or GC*.

\subsection{Fits to the UVIT photometry}

The results of our UV spectrum fits are given in  Table~\ref{tab:MatchBBfit}.
For the 10 foreground stars (source number with (fg)), the radius in column 6 is calculated using the Gaia distance. 
We compare the $\chi^2$ values for the BB and PL fits to find if the fits are statistically good
(marked Bestfit in the last column). 
To label a model a bestfit model, we require $\chi^2<3$ for at least one of BB or PL.
If both models have $\chi^2<3$ and the fit with the higher $\chi^2$ has $\chi^2$
less than the $\chi^2$ of the bestfit plus 1, we mark both PL and BB as bestfit models.
For 11 of the 49 sources in  Table~\ref{tab:MatchBBfit}, neither BB nor PL gives a good fit to the UV spectrum. 
For 29 sources, both BB and PL give a good fit. 
For 6 sources only BB gives a good fit and for 3 sources  only PL gives a good fit.

\subsubsection{Blackbody fits}

Figure~\ref{fig:RvsT} shows $T_{eff}$, $R/R_{\odot}$ and 90\% errors 
from the  blackbody fits to the UVIT photometry, excluding the sources identified as foreground stars. 
Evolutionary tracks\footnote{These tracks are from simple stellar models constructed 
using  EZ-Web at astro.wisc.edu/~townsend.} for stars of mass 10, 20, 30, 50 and 100 $M_{\odot}$ 
are plotted for comparison.
The good fits, with $\chi^2<3$, are marked with blue symbols. Comparison of the good fits to the models
shows consistency with evolved stars with mass between 10 and 30 $M_{\odot}$.
The six sources with fit radii above the 100 $M_{\odot}$ evolutionary track could be
compact clusters of massive stars. E.g., a hundred 30 $M_{\odot}$ stars of 
radius 500 $R_{\odot}$ have an effective radius 10 times that of a single 30 $M_{\odot}$ 
consistent with the fits. 

\subsubsection{Powerlaw fits}

Table \ref{tab:MatchIndices} lists both X-ray PL index ($\alpha_{X}$) and UV to X-ray  PL index 
($\alpha_{UVX}$). Table~\ref{tab:MatchBBfit}  lists UV PL index ($\alpha_{UV}$).
Figure~\ref{fig:indexdist} shows the distribution of X-ray indices, UV to X-ray indices
and of UV indices.
The X-ray indices span a wide range, $\sim$-1.5 to 5.5. The most negative index expected is -2,
corresponding to the Rayleigh-Jeans long wavelength side of a blackbody, and high values can
be from the Wien side (high energy tail) of a blackbody. 
Nonthermal spectra have expected indices in the range 0 to 2. 
The UV to X-ray indices cover a narrow range, $\sim$0.5 to 2.5, roughly consistent with 
standard non-thermal spectra from UV to X-ray. 
The UV indices cover a wide range, $\sim$0 to 9. The ones from 0 to 2 are consistent with standard
non-thermal spectra. The ones that are higher likely indicate the Wien tail of a thermal spectrum
which has temperature lower than the longest wavelength UV band observed for that source. 
E.g. if measured in NUVN2 filter, the source has a temperature less than $\sim$10,000 K.
None of the UV or UV to X-ray indices is $<0$, showing that from near UV to far UV and from
far UV to soft X-ray, the spectra are all decreasing with increasing energy.
The sources with  X-ray index $<0$ are likely caused by a hard X-ray excess, as expected
for HMXBs.

\subsection{Analysis of broadband UV and X-ray photometry}

Figure~\ref{fig:indexcompare} shows X-ray index and UV index vs UV to X-ray index for the sources.
The line of agreement of indices is shown as the dashed black line. 
Agreement of $\alpha_{X}$ and $\alpha_{UVX}$ shows that the spectrum from X-ray to UV 
(excluding the low energy UV bands) is consistent with a single PL. 
Of the 59 sources both indices, $\alpha_{X}$ and $\alpha_{UVX}$ agree within 90\% errors 
for 16 sources.
Thus for $\sim$1/4 of the sources the indices match well;  
for  $\sim$1/4 of the sources the indices are consistent within 3$\sigma$; 
and for the remaining  $\sim$1/2, the indices are significantly different.
Similarly, agreement of $\alpha_{UVX}$ and $\alpha_{UV}$ shows that the spectrum from 
X-ray (excluding the high energy X-ray band) to UV is consistent with a single PL. 
About 1/3 sources (10 of 31) show agreement within the 90\% errors.
In addition, there are several cases
where $\alpha_{UV}$ vs. $\alpha_{UVX}$ agree to within the 3$\sigma$ errors.

The PL index falls in the range of 1 to 3 for sources with consistent $\alpha_{X}$ and $\alpha_{UVX}$
or $\alpha_{UV}$ and $\alpha_{UVX}$. 
For all 22 cases with steep UV PL index ($>$3), the BB fit is better than 
or as good as the PL fit. 
This confirms that the spectrum is most likely a thermal  spectrum.
The temperatures for these BB fits are all below $\simeq$12000 K, and steeper values of
index are related with lower temperatures, further confirming that the spectra are BB or stellar.

The line of X-ray PL index $=-1/3$  is marked on Figure~\ref{fig:indexcompare}, which corresponds to
the spectrum expected for a multi-temperature accretion disk \citep{1986ApJ...308..635M}.
This is the spectrum expected for an LMXB which is dominated by emission from the accretion disk
if the inner disk temperature is as high as the hard X-ray band (2-8 keV) used for the X-ray catalog 
\citep{2016MNRAS.461.3443V}.
From Table 1, we find 8 of the sources (nos. 2, 27, 32, 40, 46, 49, 57 and 60)  have X-ray index 
consistent or near this value.
Of these 8, only nos. 27 and 57 were previously identified as LMXBs. 
Most identified LMXBs have X-ray PL index between 1 and 2. 
This can be explained if the inner disk temperature is lower than 2 keV, which is expected
for many LMXBs.
None of these 8 sources has  $\alpha_{UVX}$ consistent with -1/3, which indicates that the X-ray to 
UV spectrum is not described by a disk blackbody. This could be caused by several reasons, including
an outer disk temperature hotter than the shortest UV wavelength (temperature of 20,000 K) or
a companion star which contributes to the UV fluxes.

\section{Discussion}

Here we study the properties of X-ray sources associated with UV sources in M31 made possible by the high spatial resolution of UVIT.
The most numerous identified X-ray type is LMXB, with 15 sources. 
Here (Table~\ref{tab:MatchIndices}) we found that all 15 LMXBs which are detected with UVIT are associated with globular clusters.
This is not surprising because the expected UV emission from an LMXB at the distance of M31 is below the sensitivity limit of UVIT, 
whereas the UV emission from BHB stars in a globular cluster in M31 is readily detectable. 

9 of those were identified with stellar clusters (using \citealt{2015ApJ...802..127J}).
Their synthetic cluster analysis (Fig. 10 in \citealt{2015ApJ...802..127J}) shows that the
detection percentage for clusters as a function of cluster age is roughly constant as a function of log(age). 
Thus a significant number ($\sim1/4$) of their cluster IDs should 
correspond to old clusters, with age $>10^9$ yr.
The UVIT/X-ray LMXB sources which have stellar cluster identifications from \citet{2015ApJ...802..127J}  are plotted in
a color-color diagram in the left panel of Figure~\ref{fig:clusters} as red points.
We find that the UVIT/X-ray/LMXB/cluster sources are all located in the region where luminous globular
clusters reside \citep{2015ApJ...802..127J}.
This is also seen in the color-magnitude diagram in the right panel of  Figure~\ref{fig:clusters}:
the UVIT/Chandra/LMXB/cluster sources reside in some of the most luminous globular clusters in M31.

In  Figure~\ref{fig:RvsT} the LMXBs are marked by black squares. The LMXBs make up
many of the  UVIT/Chandra sources in the cool $T_{eff}$ part of the $R$ vs. $T_{eff}$ plot. 
The UVIT emission  associated with LMXBs is properly interpreted as globular cluster emission rather than as single evolved massive stars.
In particular, blue horizontal branch stars (BHB) have luminosity and effective temperatures and luminosities, that depends on stellar mass, which are  
$\sim$100 to 2000 L$_{\odot}$ and $\sim$5,000 to 12,000 K, for masses up to 5 M$_{\odot}$.  
The horizontal branch lifetime is $\sim$10-30\% of the main sequence lifetime. Thus a globular cluster with $10^6$ stars with 1\% BHB
will have a stellar BHB luminosity $\sim 10^6$ L$_{\odot}$, comparable to what is implied by the blackbody fits to the UVIT photometry.
I.e. the UVIT luminosity is dominated by BHB stars for the globular cluster associated LMXBs.

The next most common X-ray type is AGN, with 5 in Table~\ref{tab:MatchIndices}. 
These have X-ray PL index between 1 and 2. 
The UVIT photometry for No.30 is consistent with PL (index 1.7) or hot star  (T$\sim$17 K); 
for No.36, it is  more consistent with PL (index 0.9) than hot star (T$\sim$21K);
for No.39 
and No.43, it is not consistent with either PL or hot star; 
for No.55, it is consistent with both PL (index 0.5) or hot star (T$\sim$19K).
As noted above, 
hot star spectra and galaxy/AGN spectra
are not distinguishable for sources with only 3 bands of UVIT photometry.

 Many sources  in Table~\ref{tab:MatchIndices} had no previous X-ray identification in \citet{2016MNRAS.461.3443V} or 
 optical identification in \citet{2015ApJ...802..127J}. 
 We expanded the number of identifications by searching the recent Gaia DR2 catalogs and newer catalogs of sources such as globular clusters in M31.
Of the 67 UVIT/Chandra sources, 9 are found to be foreground stars (separation $<1^{\prime\prime}$)
and 2 more are found where the UVIT sources is likely a foreground star (separation $1-2^{\prime\prime}$).
20 are found to be associated with globular clusters ($<1^{\prime\prime}$) and 3 more where the UVIT sources is likely a globular cluster ($1-2^{\prime\prime}$).

The SNR is source No.34, with steep X-ray PL index indicating a thermal X-ray spectrum.
Neither BB nor PL fits the UV spectrum, thus the SNR UV emission could be dominated by UV 
emission lines like Galactic SNRs. 

\section{Conclusion}\label{sec:summary}

\begin{figure}
\plotone{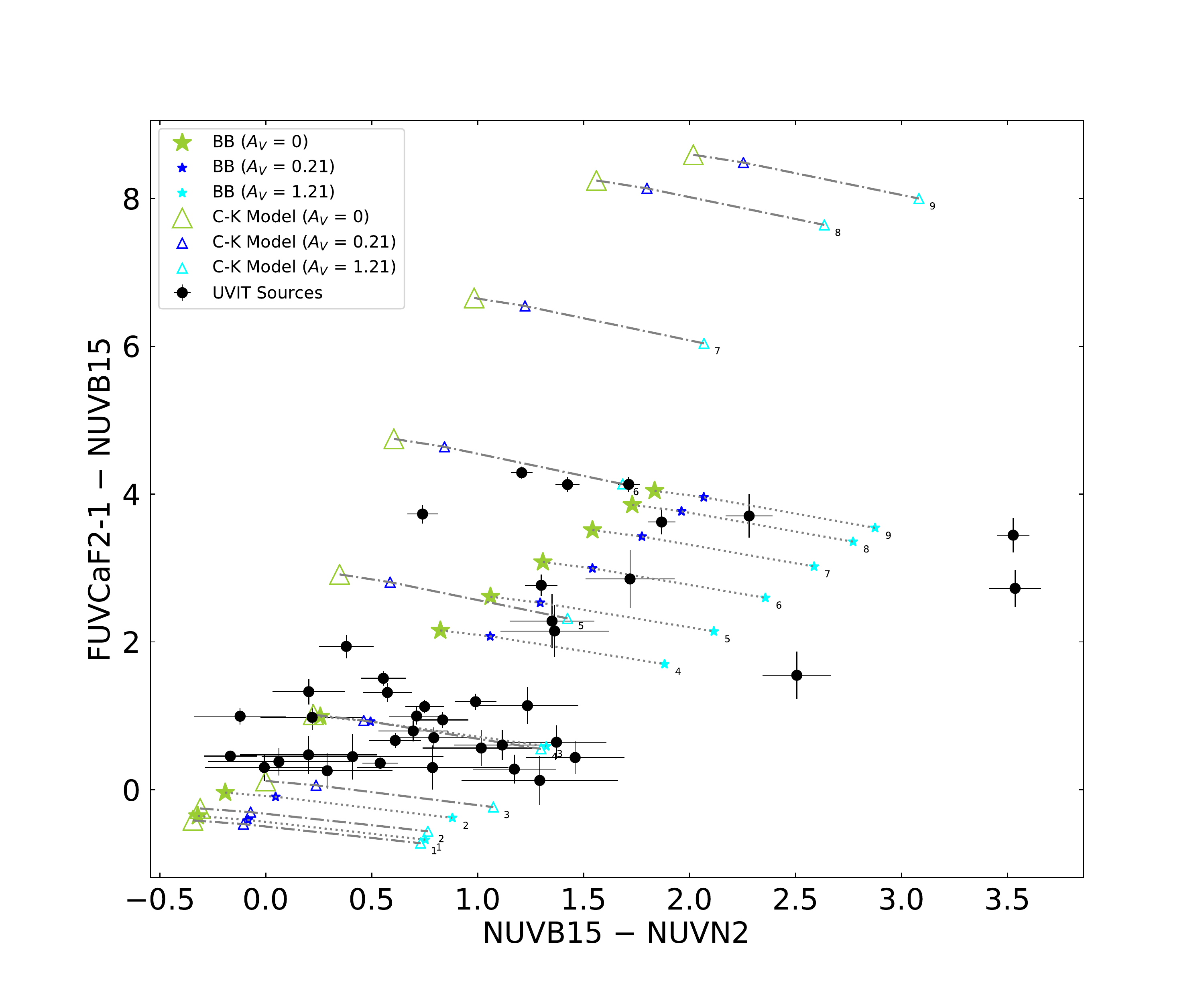}
\caption{UV Color-color diagram with FUVCaF2 $-$ NUVB15 vs. NUVB15 $-$ NUVN2. The data are plotted as black circles with error bars. 
Model points are calculated for a series of black bodies (star symbols) and stellar models (triangles, labelled C-K \citealt{2004Castelli}). The 
labels, temperatures and corresponding spectral types are: 1  45000K O5V; 2  30000K B0V; 3 15000K B5V; 4 9500K A0V; 
5 8200K A5V; 6 7200K F0V; 7  6400K F5V; 8  6000K G0V; 9 5700K  G5V. 
Extinctions of $A_V = 0$ (green), $A_V = 0.21$ (blue) and $A_V = 1.21$ (cyan) are plotted for each model, connected by dotted lines.}
\label{fig:colorcolordiag1}
\end{figure}

\begin{figure}
\plotone{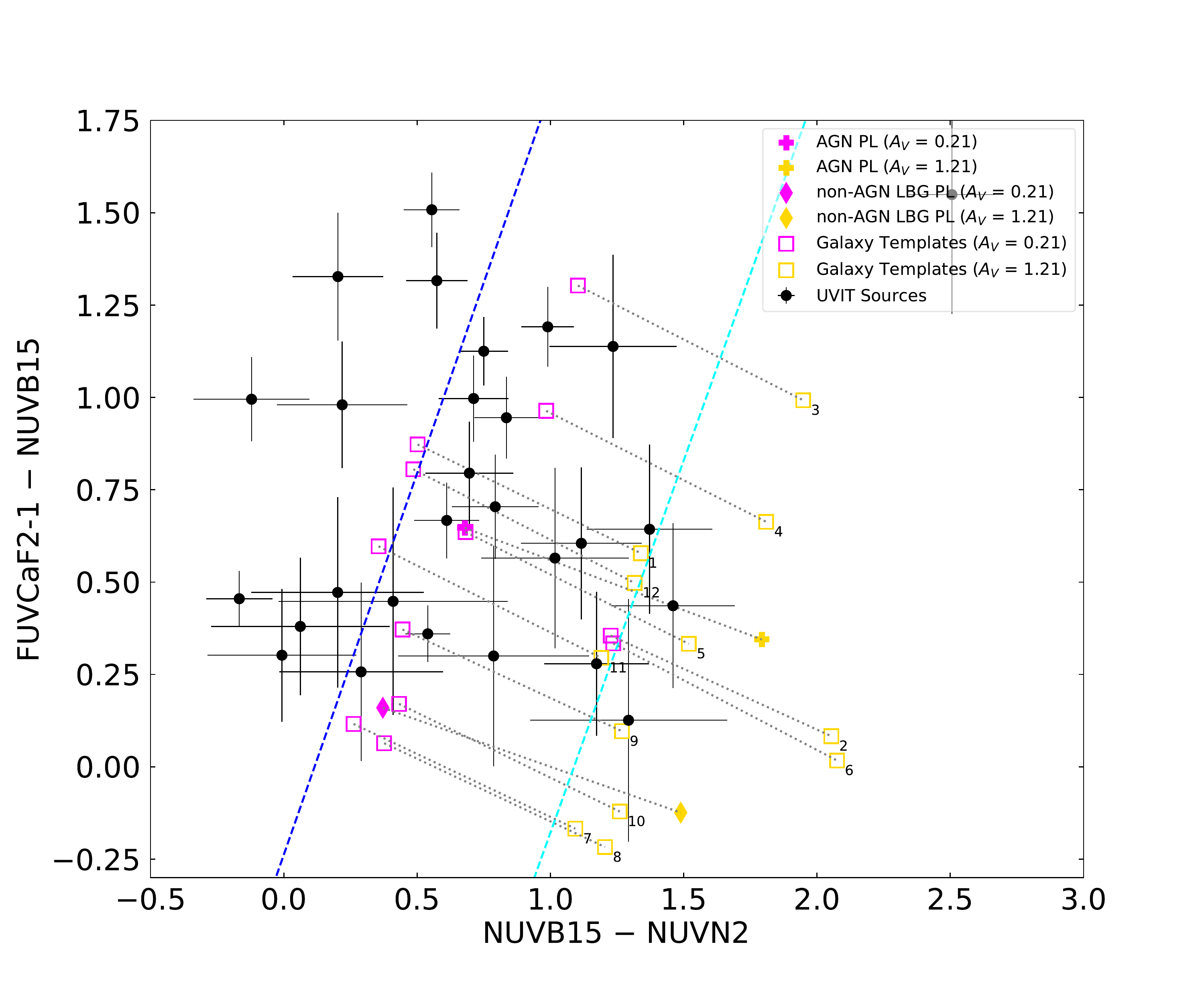}
\caption{UV Color-color diagram with FUVCaF2 $-$ NUVB15 vs. NUVB15 $-$ NUVN2 for the lower left
region of the previous figure. 
The calculated points  assuming the counterparts of the X-ray sources are background galaxies behind M31 are plotted. The series of galaxy spectrum templates are from the Kinney-Calzetti Spectal Atlas of Galaxies at STScI.
For AGNs and non-AGN Lyman-break galaxies \citep{2011ApJ...733...31H}, the spectrum is a power law $F_{\lambda} \propto \lambda^{\beta}$, with indices $\beta = -0.314$ and $\beta = -1.49$, respectively. 
Template galaxy spectra labeled 1-12 are: elliptical, bulge, S0, Sa, Sb, Sc, starburst with E(B-V)$<$0.1, starburst with 0.11$<$E(B-V)$<$0.21, starburst with 0.25$<$E(B-V)$<$0.35, starburst with 0.39$<$E(B-V)$<$0.50, starburst with 0.51$<$E(B-V)$<$0.60, and starburst with 0.61$<$E(B-V)$<$0.70. The blackbody models from Fig. 3 for various temperatures are connected  by the dotted lines, with 
$A_V = 0.21$ (blue line) and $A_V = 1.21$ (cyan line). For this range of FUVCaF2-NUVB15 color, the region covered by C-K stellar models is nearly the same as for blackbody models.}
\label{fig:colorcolordiag2}
\end{figure}

\begin{figure}
\plotone{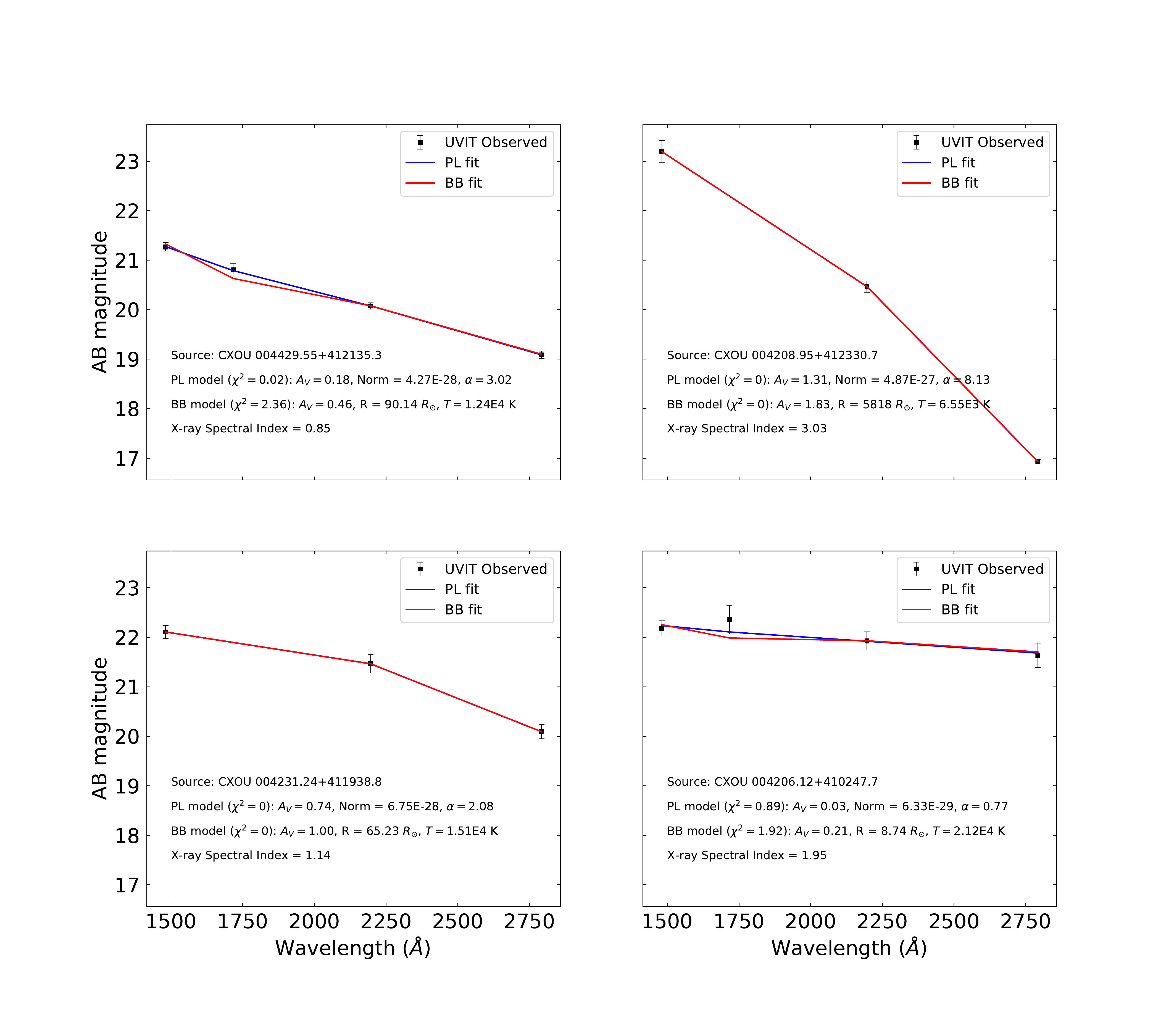}
\caption{Sample spectral fits to the UVIT photometry.
}
\label{fig:spectra}
\end{figure}

\begin{figure}
\plotone{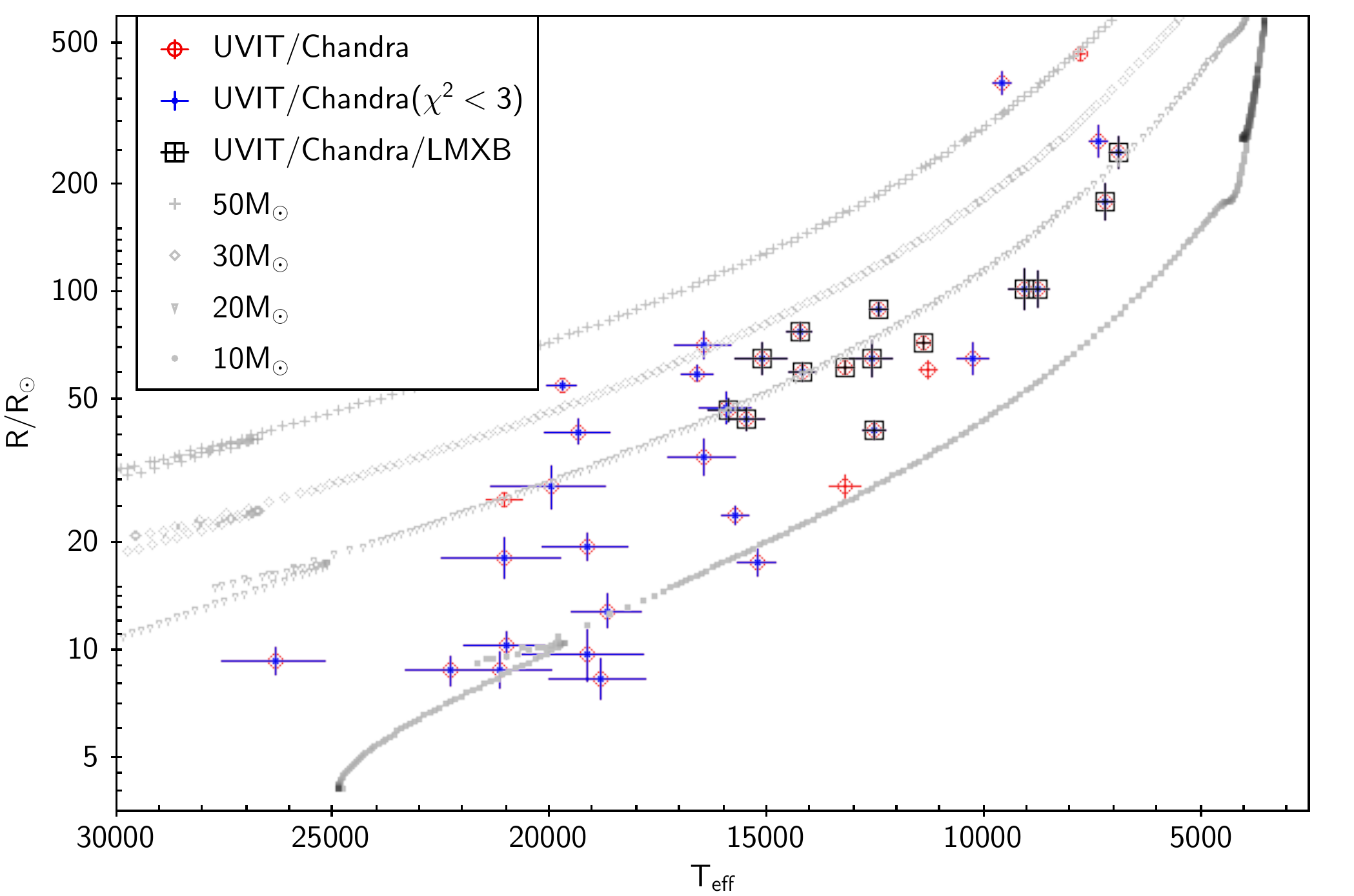}
\caption{$T_{eff}$ and $R/R_{\odot}$ from the blackbody fits to the UVIT photometry of the
UVIT/Chandra sources (red circles with error bars). Foreground stars are omitted. The sources identified with LMXBs are marked with 
 black squares. Sources for which the blackbody fits are good
($\chi^2<3$) are shown with additional blue markers. Evolutionary tracks for star of mass 10,
20, 30, and 50 $M_{\odot}$ are shown for comparison: the evolution is generally from left to right
(decreasing $T_{eff}$ and increasing $R$). 
}
\label{fig:RvsT}
\end{figure}

\begin{figure}
\plotone{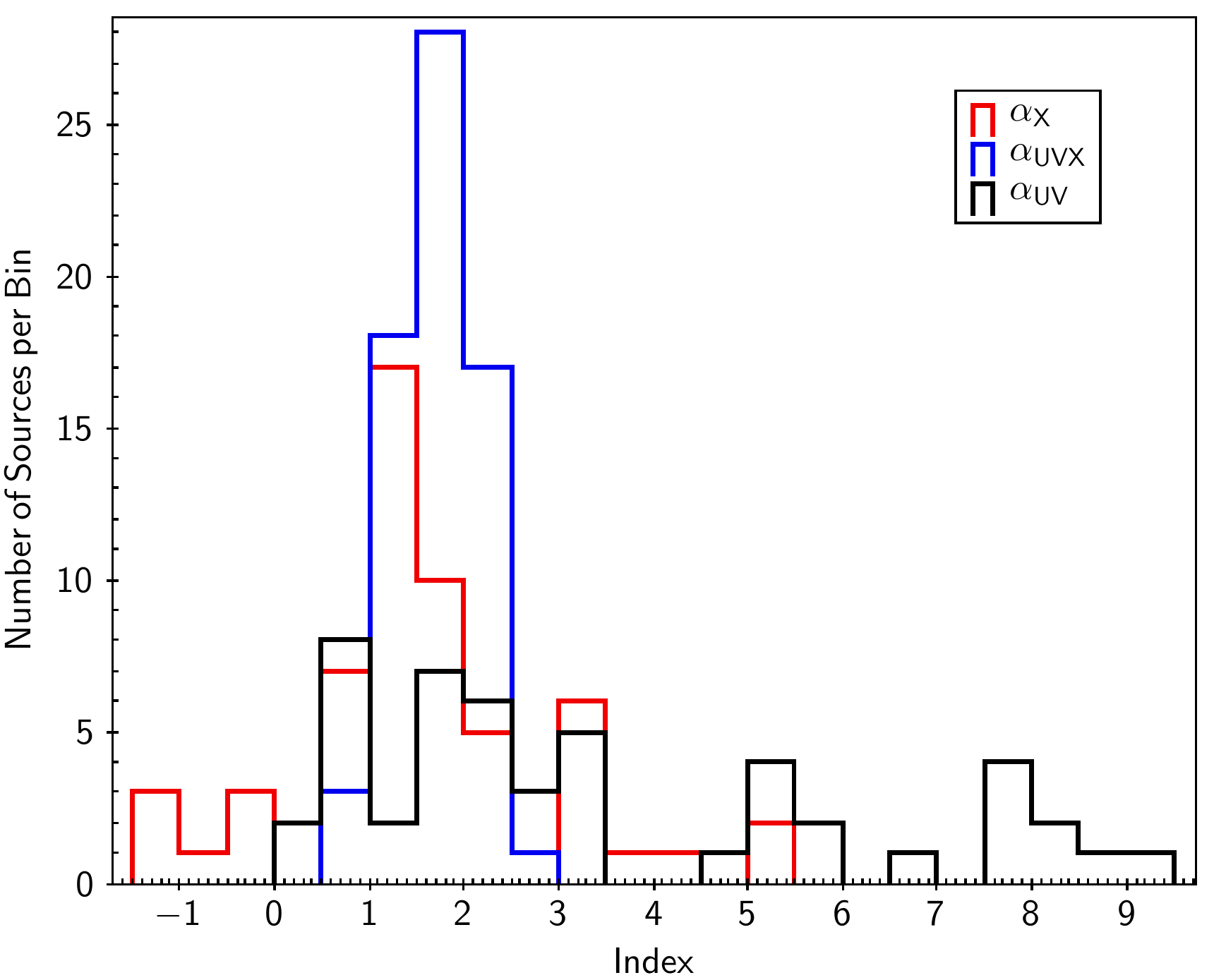}
\caption{The distributions of X-ray power-law spectral index (red histogram), UV to X-ray 
 index (blue histogram), and UV index (black histogram). For X-ray index and UV to X-ray index the fits were to 2 bands, so all sources with data are included. 
  For the UV index, sources which had poor power-law fits ($\chi^2>3$) were excluded.
}
\label{fig:indexdist}
\end{figure}

We utilize the new UVIT catalog of M31 
to find matches with the Chandra catalog of M31, and carry out a study of the nature of the
sources. The UV color-color diagram shows consistency with stellar 
UV colors for a range of main sequence stars (Fig.~\ref{fig:colorcolordiag1}), or with galaxies and AGN
 (Fig.~\ref{fig:colorcolordiag2}). 
Spectral fits to the UVIT photometry with BB and PL models were carried out.
The BB models give good fits for 35 cases,  
consistent with evolved stars 
with mass between $\sim 10$ and 30 $M_{\odot}$ (Fig.~\ref{fig:RvsT}). 
Thus a number of our sources are likely associated with massive stars, 
excepting those associated with globular clusters as discussed below.

\begin{figure}
\plotone{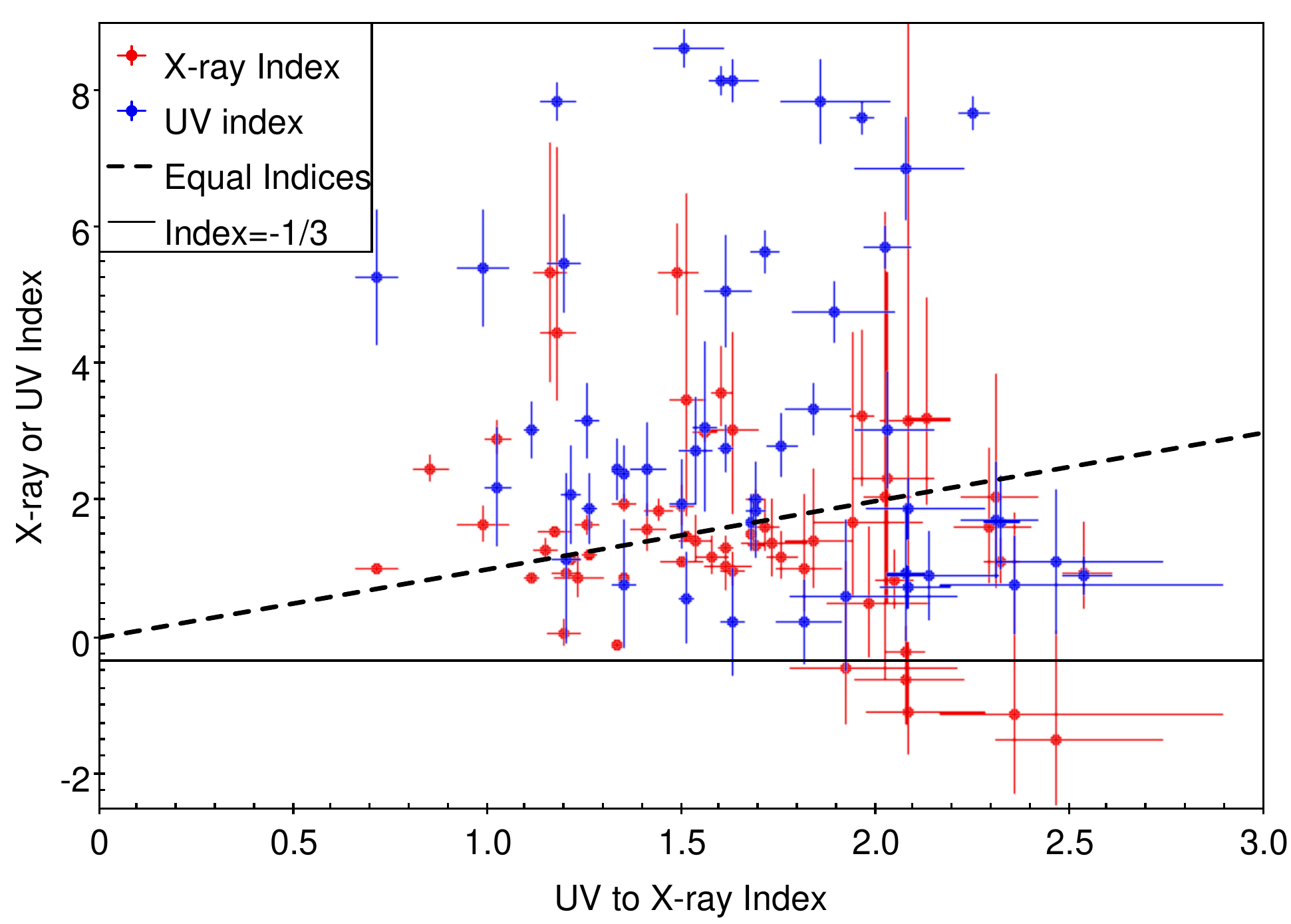}
\caption{X-ray index (red circles) and UV index (blue circles) vs UV to X-ray index.
The line of equal indices is shown as a dashed black line and the line of index of $-1/3$ is shown as a black solid line. 
The latter line corresponds to the expected slope of a multi-temperature accretion disk 
 \citep{1986ApJ...308..635M} with 
$kT_{in}$ higher than the highest energy band and with  
$kT_{out}$ lower than the lowest energy band.
}
\label{fig:indexcompare}
\end{figure}

\begin{figure}
\plottwo{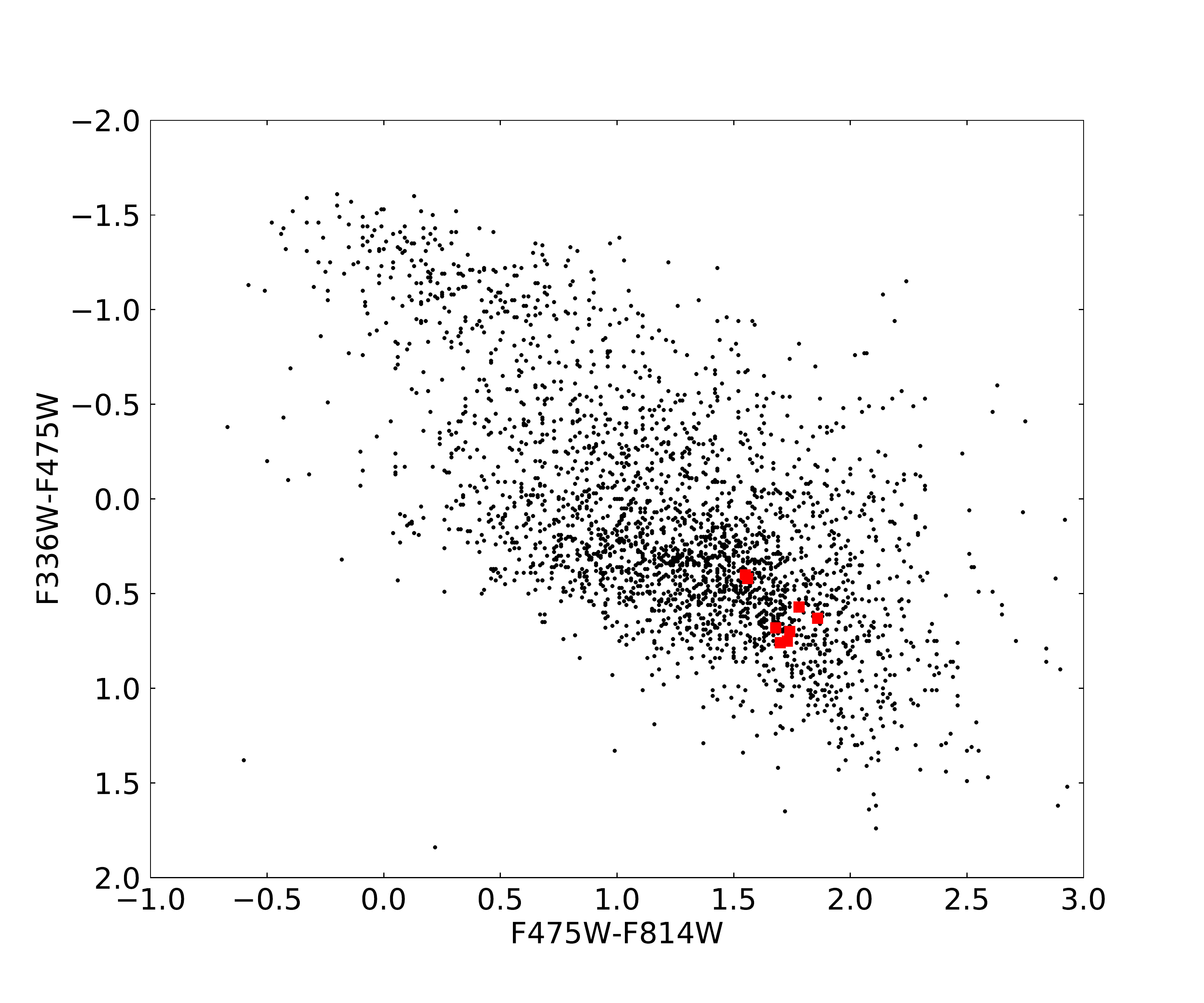}{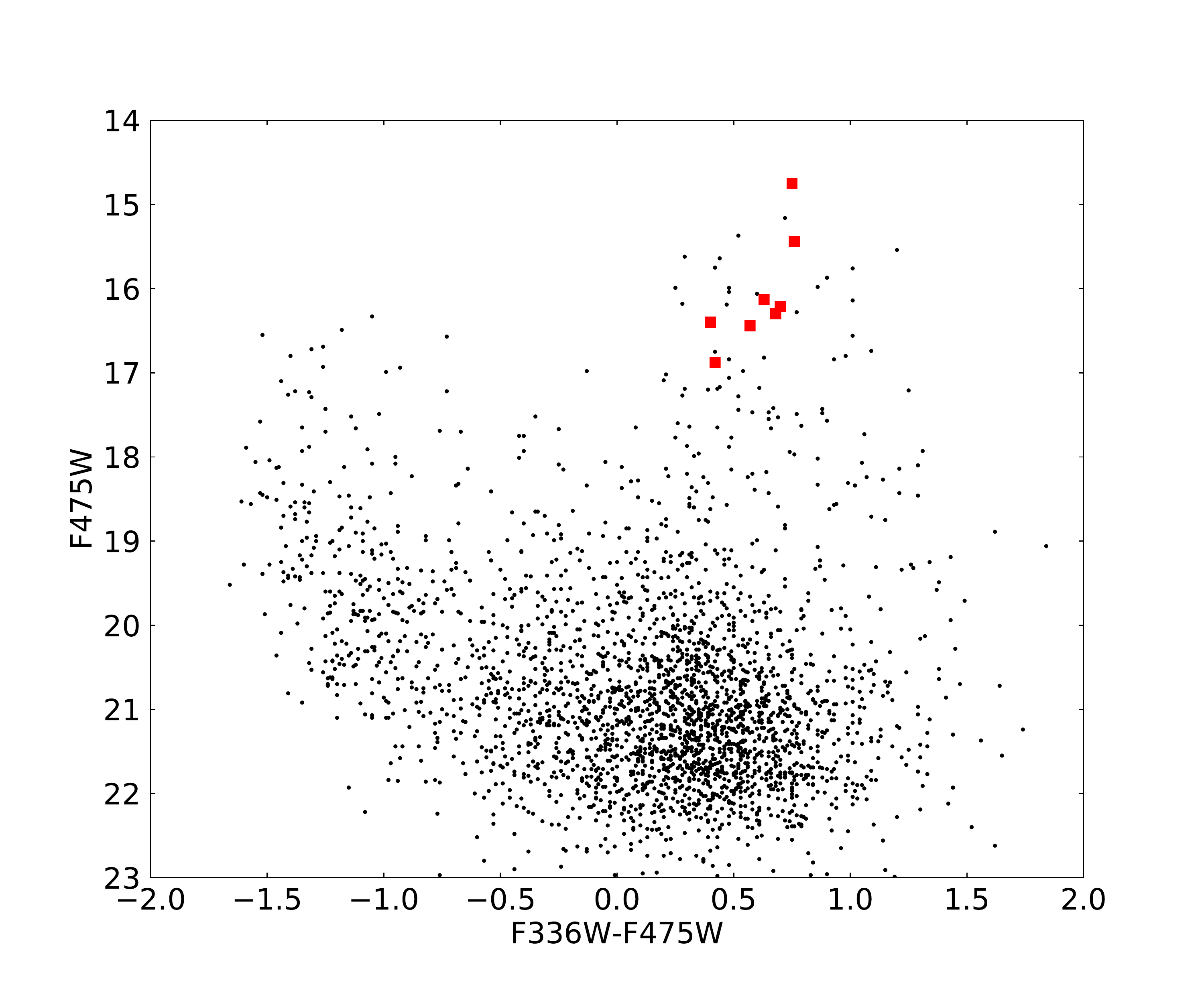}
\caption{Left: 
Color-color diagram of  M31 clusters from \citet{2015ApJ...802..127J} (black points) and those matched with the UVIT/LMXBs in Table~\ref{tab:MatchIndices} (red points).  Right: color magnitude diagram
of M31 clusters (black points) and clusters matched with the UVIT/LMXBs (red points), showing that the
UVIT/LMXBs are located in luminous globular clusters.
}
\label{fig:clusters}
\end{figure}

The PL models give good fits to the UVIT photometry for 32 sources, and are an alternate description.
However, all cases with PL index $>3$ are fit better with BB models. 
We derive PL indices from the two-bands of X-ray photometry, yielding $\alpha_{X}$, 
and from the hardest UV band and softest X-ray band, yielding $\alpha_{UVX}$.
Comparing  $\alpha_{UV}$, $\alpha_{X}$ and $\alpha_{UVX}$  (Fig.~\ref{fig:indexcompare}),
we find approximately 3 groups of sources: those with equal indices in different bands (lying near the
line of equal indices); those with steep UV index which are better fit with BB spectra; 
and a small group with $\alpha_{X}\sim-1/3$, consistent with accretion disk spectra. 
The latter group has  $\alpha_{UVX}$ between 1 and 2.5, showing that the accretion disk spectrum 
does not extend down to the UV band.

Cross-matches was carried out using the VizieR tool to catalogs of stars, clusters and other source types.
We matched up to separation $2^{\prime\prime}$ to include all possible matches to the UVIT sources.
The X-ray source position errors are smaller, so only matches $<1^{\prime\prime}$ are likely matches to the X-ray sources.
20 of the UVIT/Chandra sources match with M31 globular clusters and 9 with foreground stars to $<1^{\prime\prime}$.
3 more globular clusters and 2 more foreground stars are consistent with the UVIT source positions with separation $1-2^{\prime\prime}$. 
The UV emission of the UVIT/Chandra sources associated with globular clusters is consistent with emission from blue horizontal branch stars rather than from the X-ray source.
The LMXBs associated with globular clusters are among the most luminous globular clusters in M31.

The UVIT/Chandra sources are generally faint in UV (AB magnitudes $\sim20$ to 23), 
thus deeper UV photometry is needed to better study these sources
and more accurate positions are needed to verify source identifications in optical and X-ray.
Future work includes obtaining better photometry and more accurate positions for sources in M31 in the UVIT bands. 

\acknowledgments
This project is undertaken with the financial support of the Canadian Space Agency and
from the Natural Sciences and Engineering Research Council of Canada.
This research has made use of the VizieR catalogue access tool, CDS, Strasbourg, France (DOI: 10.26093/cds/vizier). 
The original description of the VizieR service was published in \citet{2000AAS..143...23O}.
We thank the referee for recommendations which led to improvements in the paper.

 \clearpage
 
\clearpage
 
\clearpage
 
\begin{longrotatetable}
\begin{deluxetable*}{cccccccccccccc}
\tablecaption{UVIT sources matched with Chandra sources, and power law indices\label{tab:MatchIndices}}
\tablewidth{800pt}
\tabletypesize{\scriptsize}
\tablehead{\colhead{No.} &
\colhead{CXOU} & \colhead{RA} & 
\colhead{Dec} & \colhead{F148W} & 
\colhead{F169M} & \colhead{F172M} & 
\colhead{N219M} & \colhead{N279N} & 
\colhead{X-ray} & \colhead{UV-to-X-ray} & 
\colhead{X-ray} & \colhead{Optical}  \\
 & & \colhead{(deg)} & \colhead{(deg)} 
 & \colhead{$m_{AB}$} & 
\colhead{$m_{AB}$} & \colhead{$m_{AB}$} & 
\colhead{$m_{AB}$} & \colhead{$m_{AB}$} & 
\colhead{index$^{a}$} & \colhead{index$^{b}$} & 
\colhead{Class$^{c}$} & \colhead{Ctpt$^{d}$}  
} 
\startdata
1 &004344.61+412409.9 & 10.9359 & 41.4028 & 22.99(.13)* &          &          &          &          & $1.28^{(0.17)}_{(0.15)}$  & $1.15^{(0.03)}_{(0.03)}$ & XRB 2E161$^{v}$          &          \\
2 &004342.68+412606.8 & 10.9279 & 41.4352 & 21.18(.06) &          &          & 20.19(.10) & 20.31(.19)* & $-1.11^{(1.03)}_{(0.60)}$ & $2.09^{(0.19)}_{(0.11)}$ &                 &          \\
3 &004233.13+410328.8 & 10.6381 & 41.0580 & 21.21(.07) &          & 20.91(.14) & 20.09(.06) & 19.34(.07) & $1.30^{(0.13)}_{(0.12)}$  & $1.62^{(0.02)}_{(0.02)}$ & LMXB         &     GC     \\
4 &004231.29+410437.1 & 10.6304 & 41.0770 & 22.09(.09)* &          & 21.62(.20) & 21.11(.14)* & 20.89(.20) & $1.92^{(0.32)}_{(0.27)}$  & $1.50^{(0.03)}_{(0.03)}$ &          &          \\
5 &004429.55+412135.3 & 11.1231 & 41.3598 & 21.27(.09) &          & 20.81(.13) & 20.08(.06) & 19.09(.08) & $0.85^{(0.06)}_{(0.06)}$  & $1.12^{(0.02)}_{(0.02)}$ & LMXB, XRB$^{v}$ & C, c, GC   \\
6 &004234.38+405709.1 & 10.6433 & 40.9525 & 21.91(.10) &          & 21.40(.13) & 20.59(.08) & 20.02(.09) & $1.17^{(0.37)}_{(0.30)}$  & $1.76^{(0.04)}_{(0.04)}$ & LMXB     &  GC        \\
7 &004152.83+404709.9 & 10.4702 & 40.7861 & 21.11(.07) & 21.11(.09) & 20.87(.14) & 19.60(.07)* & 19.05(.08) & $1.42^{(1.04)}_{(0.68)}$  & $1.84^{(0.10)}_{(0.07)}$ & LMXB &  GC* \\                
8 &004311.64+410605.7 & 10.7985 & 41.1016 & 23.03(.18)* &          &          &          &          & $0.49^{(1.11)}_{(0.77)}$  & $1.99^{(0.16)}_{(0.11)}$ &                                         &          \\
9 &004316.36+411630.0 & 10.8182 & 41.2750 & 20.89(.07) &          &          & 16.60(.04) & 15.40(.04) &                         & $2.25^{(0.04)}_{(0.04)}$ &                                         &     GDfg      \\
10 &004220.45+412639.8 & 10.5852 & 41.4444 & 23.69(.25)* &          &          &          &          & $0.87^{(0.34)}_{(0.28)}$  & $1.24^{(0.06)}_{(0.06)}$ &                                         &        \\
11 &004245.93+411036.3 & 10.6914 & 41.1768 & 23.37(.23)* &          &          &          &          & $1.55^{(0.03)}_{(0.03)}$  & $1.18^{(0.04)}_{(0.04)}$ &                                         &          \\
12 &004208.95+412330.7 & 10.5373 & 41.3919 & 23.19(.23)* &          &          & 20.47(.12)* & 16.93(.04) & $3.03^{(1.43)}_{(1.23)}$  & $1.63^{(0.07)}_{(0.06)}$ &         &    GDfg      \\
13 &004241.08+411101.7 & 10.6712 & 41.1838 & 24.21(.34)* &          &          &          &          & $1.36^{(0.66)}_{(0.46)}$  & $1.74^{(0.09)}_{(0.08)}$ &                                         &          \\
14 &004253.64+412551.0 & 10.7235 & 41.4308 & 23.14(.22) &          &          &          &          & $5.34^{(1.89)}_{(1.60)}$  & $1.16^{(0.04)}_{(0.04)}$ &                                         &          \\
15 &004240.89+412216.2 & 10.6704 & 41.3712 & 21.57(.10) &          &          & 17.44(.04)* & 16.02(.04) & $3.22^{(1.24)}_{(1.01)}$  & $1.97^{(0.03)}_{(0.03)}$ &                  &   GDfg       \\
16 &004321.06+411750.5 & 10.8378 & 41.2974 & 21.48(.09) &          &          & 21.18(.15) & 21.19(.23) & $1.48^{(0.04)}_{(0.04)}$  & $1.51^{(0.02)}_{(0.02)}$ &                     &          \\
17 &004243.66+411241.8 & 10.6819 & 41.2116 & 22.22(.14)* &          &          & 21.66(.20) & 20.64(.19) & $2.07^{(1.79)}_{(1.33)}$  & $2.31^{(0.11)}_{(0.09)}$ &                     & C, GC*     \\
18 &004241.43+411523.8 & 10.6727 & 41.2566 & 21.13(.09) &          &          & 20.34(.11) & 19.64(.12) & $1.33^{(0.04)}_{(0.04)}$  & $1.69^{(0.02)}_{(0.02)}$ & LMXB        & C, GC          \\
19 &004250.81+411707.3 & 10.7117 & 41.2854 & 21.78(.12) &          &          & 21.34(.19) & 19.88(.13) & $1.09^{(0.35)}_{(0.29)}$  & $2.32^{(0.05)}_{(0.04)}$ &          & C, GC          \\
20 &004236.60+411350.3 & 10.6525 & 41.2307 & 22.75(.18) &          &          & 21.57(.19) &          & $3.01^{(0.15)}_{(0.13)}$  & $1.56^{(0.03)}_{(0.03)}$ &                     &      GDfg    \\
21 &004303.86+411804.8 & 10.7661 & 41.3014 & 21.15(.09) &          &          & 20.45(.11) & 19.66(.12) & $1.20^{(0.01)}_{(0.01)}$  & $1.26^{(0.02)}_{(0.02)}$ & LMXB          & C, GC         \\
22 &004231.24+411938.8 & 10.6302 & 41.3275 & 22.11(.13) &          &          & 21.47(.19) & 20.09(.14) & $1.15^{(0.02)}_{(0.02)}$  & $1.21^{(0.02)}_{(0.02)}$ & LMXB         & C, GC         \\
23 &004246.08+411736.1 & 10.6920 & 41.2934 & 23.56(.29) &          &          &          &          & $1.09^{(0.07)}_{(0.07)}$  & $1.50^{(0.05)}_{(0.05)}$ &                                         & C, GC          \\
24 &004336.63+410811.8 & 10.9027 & 41.1366 & 22.63(.17) &          &          & 21.49(.18) & 20.26(.16) & $2.31^{(3.01)}_{(1.82)}$  & $2.03^{(0.12)}_{(0.08)}$ & LMXB      & C, GC          \\
25 &004240.73+411004.3 & 10.6697 & 41.1679 & 22.03(.12) &          &          & 18.30(.05)* & 17.56(.05) & $2.05^{(4.17)}_{(2.66)}$  & $2.03^{(0.06)}_{(0.05)}$ &          &     GDfg     \\
26 &004218.63+411401.7 & 10.5776 & 41.2338 & 20.64(.07) &          &          & 19.70(.09) & 18.86(.09) & $0.86^{(0.02)}_{(0.02)}$  & $1.35^{(0.01)}_{(0.01)}$ & LMXB         & C, GC          \\
27 &004314.36+410721.1 & 10.8099 & 41.1225 & 20.91(.08) &          &          & 19.91(.09) & 19.20(.10) & $-0.11^{(0.02)}_{(0.02)}$ & $1.33^{(0.01)}_{(0.01)}$ & LMXB      &    GC      \\
28 &004226.14+412551.9 & 10.6089 & 41.4311 & 22.86(.18)* &          &          & 22.41(.25)* & 22.00(.35) & $0.93^{(0.08)}_{(0.08)}$  & $1.21^{(0.03)}_{(0.03)}$ &            &          \\
29 &004353.65+411655.0 & 10.9736 & 41.2820 & 23.39(.24)* &          &          &          &          & $2.46^{(0.22)}_{(0.18)}$  & $0.86^{(0.04)}_{(0.04)}$ & \textless{}XRB\textgreater{}  & n, NGC224*          \\
30 &004244.35+411608.6 & 10.6848 & 41.2691 & 20.00(.06) &          &          & 19.33(.08) & 18.72(.09) & $1.50^{(0.03)}_{(0.02)}$  & $1.68^{(0.01)}_{(0.01)}$ & AGN          &          \\
31 &004315.41+411124.6 & 10.8142 & 41.1902 & 21.92(.12) &          &          & 21.32(.17) & 20.20(.15) & $2.03^{(0.12)}_{(0.11)}$  & $1.70^{(0.02)}_{(0.02)}$ & LMXB     & C, GC          \\
32 &004438.33+412530.3 & 11.1597 & 41.4251 & 21.64(.10)* &          & 21.20(.34) & 21.26(.16) & 21.19(.30) & $-1.13^{(2.95)}_{(1.15)}$ & $2.36^{(0.54)}_{(0.19)}$ &        &          \\
33 &004412.20+413148.2 & 11.0509 & 41.5301 & 21.41(.09)* &          &          & 21.39(.17)* &          & $1.01^{(1.07)}_{(0.62)}$  & $1.82^{(0.10)}_{(0.07)}$ & \textless{}XRB\textgreater{}              & p      \\
34 &004451.08+412904.9 & 11.2129 & 41.4847 & 19.31(.05) &          & 18.88(.12) & 18.86(.06)* & 19.02(.11) & $3.18^{(6.63)}_{(2.41)}$  & $2.09^{(0.11)}_{(0.07)}$ & \textless{}SNR\textgreater{}  & s   \\
35 &004118.65+405158.6 & 10.3277 & 40.8663 & 22.31(.16) &          & 20.93(.14) & 18.68(.05) & 16.82(.04) & $4.44^{(2.72)}_{(0.99)}$  & $1.18^{(0.05)}_{(0.04)}$ &       &   GDfg       \\
36 &004206.12+410247.7 & 10.5255 & 41.0466 & 22.18(.15) &          & 22.36(.29) & 21.93(.19) & 21.64(.24) & $1.95^{(0.12)}_{(0.11)}$  & $1.36^{(0.03)}_{(0.03)}$ & AGN     &  GC        \\
37 &004210.36+405149.2 & 10.5432 & 40.8637 & 21.78(.13)* &          & 21.10(.15) & 21.50(.15)* & 20.33(.13)* &                         & $2.14^{(0.18)}_{(0.11)}$ &        &    GD*      \\
38 &004143.44+410504.7 & 10.4310 & 41.0846 & 21.03(.09) &          & 19.30(.07) & 16.90(.04) & 15.18(.03) & $3.57^{(0.67)}_{(0.47)}$  & $1.61^{(0.02)}_{(0.02)}$ &       &      GDfg     \\
39 &004126.25+405326.0 & 10.3594 & 40.8906 & 21.98(.14) &          & 21.70(.20) & 20.65(.10)* & 20.45(.14) & $1.57^{(0.38)}_{(0.30)}$  & $1.41^{(0.05)}_{(0.04)}$ & AGN  &          \\
40 &004200.45+405942.7 & 10.5019 & 40.9952 & 22.50(.18) &          & 22.63(.31) & 22.03(.19)* & 21.83(.26) & $-1.49^{(1.58)}_{(0.93)}$ & $2.47^{(0.27)}_{(0.15)}$ &      &          \\
41 &004205.62+405713.4 & 10.5234 & 40.9537 & 19.32(.05) &          & 18.96(.07) & 18.96(.06)* & 18.42(.06) & $0.93^{(0.75)}_{(0.51)}$  & $2.54^{(0.07)}_{(0.05)}$ &      &          \\
42 &004234.98+404838.8 & 10.6458 & 40.8108 & 21.90(.14) &          & 20.86(.14) & 19.96(.08)* & 19.58(.10) & $1.64^{(0.15)}_{(0.13)}$  & $1.26^{(0.03)}_{(0.03)}$ &       &         \\
43 &004241.80+405154.6 & 10.6742 & 40.8652 & 21.34(.14) &          & 20.06(.11) & 18.57(.06) & 17.27(.05) & $1.62^{(0.39)}_{(0.33)}$  & $1.72^{(0.04)}_{(0.04)}$ & AGN    &    M32      \\
44 &003823.84+401250.0 & 9.5993  & 40.2139 & 22.35(.17)* & 22.14(.14) &          &          &          & $2.89^{(0.27)}_{(0.22)}$  & $1.03^{(0.03)}_{(0.03)}$ &            &    GD*      \\
45 &004432.46+410533.7 & 11.1353 & 41.0927 & 23.00(.22) &          & 21.08(.13) & 19.56(.07)* & 16.03(.04)* &                         & $1.51^{(0.10)}_{(0.08)}$ &            &      GDfg*     \\
46 &004410.16+411830.4 & 11.0424 & 41.3085 & 22.56(.19) &          & 21.81(.19) & 22.26(.23)* & 21.48(.28) & $-0.46^{(1.54)}_{(0.80)}$ & $1.93^{(0.29)}_{(0.14)}$ &          &          \\
47 &004356.41+412202.2 & 10.9850 & 41.3673 & 24.13(.33) &   &   & 21.85(.15) & 20.50(.13) & $1.64^{(0.28)}_{(0.23)}$  & $0.99^{(0.07)}_{(0.06)}$ & LMXB, \textless{}XRB\textgreater{}  & C, c, GC    \\
48 &004229.72+405247.8 & 10.6239 & 40.8800 & 23.54(.29) &          &          & 21.99(.15) & 19.48(.07) &                         & $1.89^{(0.15)}_{(0.11)}$ &                                         &    GD      \\
49 &004257.63+412137.6 & 10.7402 & 41.3605 & 24.29(.35) &          &          & 21.44(.18)* & 19.72(.12) & $-0.63^{(0.79)}_{(0.65)}$ & $2.08^{(0.15)}_{(0.13)}$ &            &          \\
50 &004337.27+411443.6 & 10.9053 & 41.2455 & 23.92(.29) &          &          & 21.77(.20) & 20.41(.16) & $1.01^{(0.02)}_{(0.02)}$  & $0.72^{(0.05)}_{(0.05)}$ & LMXB     & C, GC         \\
51 &004301.62+411052.7 & 10.7568 & 41.1813 & 23.81(.28) &          &          & 20.10(.10) & 17.82(.06) & $5.32^{(0.72)}_{(0.61)}$  & $1.49^{(0.05)}_{(0.05)}$ &          &      GDfg    \\
52 &004248.11+411729.5 & 10.7005 & 41.2915 &          &          &          & 22.12(.26)* &          & $1.62^{(1.16)}_{(0.82)}$  & $2.30^{(0.11)}_{(0.09)}$ &                                         &          \\
53 &004409.84+412813.9 & 11.0410 & 41.4705 &          &          &          & 21.33(.16)* & 18.97(.11)* &                         & $1.86^{(0.18)}_{(0.10)}$ & fgStar                                    &   GDfg*       \\
54 &004227.41+405936.1 & 10.6142 & 40.9934 &          &          &          & 22.21(.22) & 20.55(.14) & $1.04^{(0.43)}_{(0.33)}$  & $1.61^{(0.06)}_{(0.05)}$ & LMXB     &   GC       \\
55 &004137.85+410108.2 & 10.4077 & 41.0190 &          &          &          & 20.89(.11) & 20.49(.14) & $0.98^{(0.26)}_{(0.22)}$  & $1.63^{(0.03)}_{(0.03)}$ & AGN        &        \\ 
56 &004115.38+410102.5 & 10.3141 & 41.0174 &          &          &          & 21.50(.15) & 20.45(.14) & $1.41^{(0.36)}_{(0.29)}$  & $1.54^{(0.04)}_{(0.04)}$ &                     &          \\
57 &004301.42+413017.3 & 10.7560 & 41.5048 &          &          &          & 21.98(.20) & 20.22(.13) & $0.06^{(0.20)}_{(0.17)}$  & $1.20^{(0.04)}_{(0.04)}$ & LMXB, \textless{}XRB\textgreater{} & GC \\
58 &004616.82+414300.4 & 11.5701 & 41.7168 &          &          &          & 22.23(.22) &          & $1.85^{(0.15)}_{(0.14)}$  & $1.44^{(0.04)}_{(0.04)}$ &              &    GD      \\
59 &004229.35+405750.1 & 10.6223 & 40.9639 &          &          &          &          & 19.65(.07) &                         & $2.06^{(0.09)}_{(0.06)}$ &                                         &    GD      \\
60 &004318.45+411142.2 & 10.8269 & 41.1951 & 22.62(.15) &          &          & 22.49(.29) & 21.20(.23)* & $-0.20^{(0.25)}_{(0.22)}$ & $2.08^{(0.05)}_{(0.04)}$ &                &          \\
61 &004230.10+411841.8 & 10.6254 & 41.3116 &          &          &          &          & 20.65(.18) & $3.21^{(1.76)}_{(1.24)}$  & $2.14^{(0.06)}_{(0.05)}$ &                                         &   GC, GD      \\
62 &004248.83+411512.9 & 10.7035 & 41.2536 &          &          &          &          & 19.94(.14) &                         & $2.31^{(0.06)}_{(0.05)}$ &                                         &     GC, GD     \\
63 &004325.64+411537.4 & 10.8568 & 41.2604 &          &          &          &          & 20.73(.19) & $0.82^{(0.44)}_{(0.39)}$  & $2.05^{(0.05)}_{(0.05)}$ &                                         & C, GC         \\
64 &004431.98+412519.5 & 11.1333 & 41.4221 &          &          &          &          & 20.84(.25) & $1.69^{(2.78)}_{(1.07)}$  & $1.94^{(0.18)}_{(0.10)}$ & fgStar                                    & f, GDfg         \\
65 & 004225.04+405719.3 & 10.6044 & 40.9554 &          &          &          &          & 20.99(.18)* & $1.19^{(0.30)}_{(0.25)}$  & $1.58^{(0.04)}_{(0.04)}$ &                                         &   GC*    \\  
66 &004258.77+405901.6 & 10.7449 & 40.9838 &          &          &          &          & 18.92(.08)* &                         & $1.97^{(0.11)}_{(0.07)}$ &                                         &          \\
67 & 004321.97+405754.7 & 10.8416 & 40.9652 &          &          &          &          & 21.01(.19)* & $3.46^{(3.01)}_{(1.69)}$  & $1.51^{(0.05)}_{(0.04)}$ &                 &       GD*  \\ 
\enddata
\tablenotetext{a}{The X-ray index is calculated two X-ray bands, assuming a power-law spectrum. The X-ray fluxes are corrected for a fixed extinction of $N_H=6.66\times10^{20}$cm$^{-2}$, so the index is the source intrinsic index.}
\tablenotetext{b}{The UV to X-ray index is calculated from the shortest available wavelength UV flux and the longest available wavelength X-ray band, assuming a power-law spectrum between UV and X-ray. Both UV and X-ray fluxes are corrected for a fixed extinction of $N_H=6.66\times10^{20}$cm$^{-2}$, so the index is the source intrinsic index.}
\tablenotetext{c}{For column X-ray Class, the class type from \cite{2016MNRAS.461.3443V} is given as LMXB, AGN, or otherwise blank; the Class from \cite{2018Sasaki} is given as XRB, fgStar, and $^{v}$ is attached if the source is variable.}
\tablenotetext{d}{For column Optical Ctpt, the counterpart type from \cite{2015ApJ...802..127J} is listed as C for star cluster, or otherwise blank; the counterpart type 
from \cite{2018ApJS..239...13W} is listed as n for none, p for point/star, f for foreground, s for SNR, c for cluster; 
GD indicates that there is a Gaia DR2 parallax \citep{2018AA...616A...1G} and a distance measurement \citep{2018AJ....156...58B}. Those consistent with foreground stars are marked GDfg. 
GC indicates a globular cluster from \citet{2016ApJ...824...42C} or \citet{2016ApJ...829..116S}. }
\tablenotetext{*}{UVIT-Chandra matches that have separation 1-2$^{\prime\prime}$  have a larger probability of being accidental matches. These are marked with an asterisk in the UVIT magnitude columns. 
Matches to Gaia DR2 distance sources (GD), globular clusters (GC) and other sources with separation 1-2$^{\prime\prime}$  are marked in the optical counterpart column.}
\end{deluxetable*}
\end{longrotatetable}

\clearpage

\startlongtable
\begin{deluxetable*}{ccccccccccc}
\tablecaption{Blackbody(BB) and power-law(PL) fits to UVIT photometry\label{tab:MatchBBfit}}
\tablehead{\colhead{No.} &
\colhead{CXOU} & \colhead{UVIT filter} & 
\colhead{Model} & & 
& & &
&  \colhead{Bestfit}  \\
& & & 
 \colhead{BB} & \colhead{A$_V$\tablenotemark{$a$}} & 
 \colhead{R (R$_{\odot}$)\tablenotemark{$b$}} & \colhead{T$_{\text{eff}}$ ($10^3$ K)} &
\colhead{$\chi^2$} & \colhead{$\chi^2_{\nu}$\tablenotemark{$c$}} & \\
& & & 
 \colhead{PL} & \colhead{A$_V$\tablenotemark{$a$}} & 
 \colhead{norm ($10^{-28}$)} & \colhead{index} &
\colhead{$\chi^2$} & \colhead{$\chi^2_{\nu}$\tablenotemark{$c$}} & 
}
\startdata
2&004342.68+412606.8 & F1,F4,F5       & BB & $0.00^{(0.04)}_{(0.05)}$ & $23.8^{(1.4)}_{(1.4)}$        & $15.7^{(0.33)}_{(0.32)}$ & $2.47$   & N/A      & \checkmark \\
     &          &            & PL & $0.00^{(0.04)}_{(0.05)}$ & $2.17^{(0.26)}_{(0.24)}$      & $1.88^{(0.44)}_{(0.44)}$ & $8.62$   & N/A      &                       \\ \tableline
3&004233.13+410328.8 & F1,F3,F4,F5    & BB & $0.31^{(0.03)}_{(0.03)}$ & $61.5^{(2.7)}_{(2.6)}$        & $13.2^{(0.21)}_{(0.20)}$ & $4.26$   & $4.26$   &                       \\ 
&               &            & PL & $0.02^{(0.03)}_{(0.03)}$ & $2.73^{(0.24)}_{(0.23)}$      & $2.77^{(0.34)}_{(0.34)}$ & $0.51$   & $0.51$   & \checkmark \\ \tableline
4&004231.29+410437.1 & F1,F3,F4,F5    & BB & $0.00^{(0.06)}_{(0.06)}$ & $17.4^{(1.5)}_{(1.4)}$        & $15.2^{(0.45)}_{(0.43)}$ & $0.02$   & $0.02$   & \checkmark \\
&               &            & PL & $0.00^{(0.06)}_{(0.06)}$ & $0.97^{(0.17)}_{(0.15)}$      & $1.95^{(0.64)}_{(0.63)}$ & $2.10$   & $2.10$   &                       \\ \tableline
5&004429.55+412135.3 & F1,F3,F4,F5    & BB & $0.46^{(0.04)}_{(0.04)}$ & $90.1^{(4.3)}_{(4.1)}$        & $12.4^{(0.21)}_{(0.20)}$ & $2.36$   & $2.36$   &                       \\
  &             &            & PL & $0.18^{(0.04)}_{(0.04)}$ & $4.27^{(0.42)}_{(0.38)}$      & $3.02^{(0.40)}_{(0.40)}$ & $0.02$   & $0.02$   & \checkmark \\ \tableline
6&004234.38+405709.1 & F1,F3,F4,F5    & BB & $0.11^{(0.04)}_{(0.04)}$ & $41.1^{(2.3)}_{(2.2)}$        & $12.5^{(0.25)}_{(0.24)}$ & $0.58$   & $0.58$   & \checkmark \\
  &             &            & PL & $0.00^{(0.04)}_{(0.05)}$ & $1.50^{(0.17)}_{(0.16)}$      & $2.78^{(0.46)}_{(0.46)}$ & $2.44$   & $2.44$   &                       \\ \tableline
7&004152.83+404709.9 & F1,F2,F3,F4,F5 & BB & $0.00^{(0.03)}_{(0.04)}$ & $71.9^{(3.2)}_{(3.1)}$        & $11.4^{(0.15)}_{(0.15)}$ & $14.20$  & $7.10$   &                       \\
  &             &            & PL & $0.00^{(0.03)}_{(0.04)}$ & $3.21^{(0.29)}_{(0.27)}$      & $3.33^{(0.36)}_{(0.36)}$ & $16.29$  & $8.14$   &                       \\ \tableline
9(fg) &004316.36+411630.0 & F1,F4,F5       & BB & $0.00^{(0.01)}_{(0.04)}$ & $2.34^{(0.07)}_{(0.07)}$    & $6.3^{(0.04)}_{(0.04)}$  & $94.19$  & N/A      &                       \\
     &          &            & PL & $0.00^{(0.01)}_{(0.02)}$ & $24.72^{(1.42)}_{(1.35)}$     & $7.65^{(0.22)}_{(0.22)}$ & $339.62$ & N/A      &                       \\ \tableline
12(fg) &004208.95+412330.7 & F1,F4,F5       & BB & $1.83^{(0.05)}_{(0.05)}$ & $4.91^{(0.23)}_{(0.22)}$  & $6.6^{(0.08)}_{(0.08)}$  & $0.00$   & N/A      & \checkmark \\
     &          &            & PL & $1.31^{(0.04)}_{(0.04)}$ & $48.69^{(4.75)}_{(4.33)}$     & $8.13^{(0.30)}_{(0.30)}$ & $0.00$   & N/A      & \checkmark \\ \tableline
15(fg) &004240.89+412216.2 & F1,F4,F5       & BB & $0.00^{(0.02)}_{(0.04)}$ & $1.96^{(0.06)}_{(0.06)}$    & $6.2^{(0.04)}_{(0.04)}$  & $25.22$  & N/A      &                       \\
     &          &            & PL & $0.00^{(0.01)}_{(0.05)}$ & $13.04^{(0.84)}_{(0.79)}$     & $7.60^{(0.24)}_{(0.24)}$ & $155.15$ & N/A      &                       \\ \tableline
16&004321.06+411750.5 & F1,F4,F5       & BB & $0.03^{(0.07)}_{(0.07)}$ & $8.8^{(0.8)}_{(0.8)}$         & $22.3^{(1.03)}_{(0.95)}$ & $0.00$   & N/A      & \checkmark \\
    &           &            & PL & $0.00^{(0.06)}_{(0.07)}$ & $1.10^{(0.21)}_{(0.18)}$      & $0.55^{(0.66)}_{(0.65)}$ & $0.32$   & N/A      &    \checkmark    \\ \tableline
17&004243.66+411241.8 & F1,F4,F5       & BB & $0.75^{(0.09)}_{(0.09)}$ & $34.6^{(4.2)}_{(3.7)}$        & $16.5^{(0.81)}_{(0.74)}$ & $0.00$   & N/A      & \checkmark \\
    &           &            & PL & $0.50^{(0.09)}_{(0.09)}$ & $2.98^{(0.75)}_{(0.60)}$      & $1.70^{(0.85)}_{(0.84)}$ & $0.00$   & N/A      & \checkmark \\ \tableline
18&004241.43+411523.8 & F1,F4,F5       & BB & $0.41^{(0.05)}_{(0.05)}$ & $44.2^{(3.2)}_{(3.0)}$        & $15.5^{(0.43)}_{(0.41)}$ & $0.00$   & N/A      & \checkmark \\
   &            &            & PL & $0.15^{(0.05)}_{(0.05)}$ & $3.43^{(0.50)}_{(0.44)}$      & $2.01^{(0.53)}_{(0.53)}$ & $0.00$   & N/A      & \checkmark \\ \tableline
19&004250.81+411707.3 & F1,F4,F5       & BB & $1.14^{(0.08)}_{(0.08)}$ & $71.6^{(7.0)}_{(6.3)}$        & $16.4^{(0.67)}_{(0.63)}$ & $0.00$   & N/A      & \checkmark \\
      &         &            & PL & $0.90^{(0.07)}_{(0.07)}$ & $12.85^{(2.60)}_{(2.17)}$     & $1.69^{(0.67)}_{(0.67)}$ & $0.00$   & N/A      & \checkmark \\ \tableline
20(fg) &004236.60+411350.3 & F1,F4          & BB & $0.30$                   & $6.98^{(0.14)}_{(0.08)}\times10^{-3}$        & $13.2^{(0.57)}_{(0.53)}$ & $0.00$   & N/A      & \checkmark \\
     &          &            & PL & $0.30$                   & $1.55^{(0.46)}_{(0.35)}$      & $3.06^{(1.26)}_{(1.22)}$ & $0.00$   & N/A      & \checkmark \\ \tableline
21&004303.86+411804.8 & F1,F4,F5       & BB & $0.52^{(0.05)}_{(0.05)}$ & $46.4^{(3.3)}_{(3.1)}$        & $15.9^{(0.45)}_{(0.43)}$ & $0.00$   & N/A      & \checkmark \\
     &          &            & PL & $0.27^{(0.05)}_{(0.05)}$ & $4.43^{(0.65)}_{(0.57)}$      & $1.87^{(0.52)}_{(0.52)}$ & $0.00$   & N/A      & \checkmark \\ \tableline
22&004231.24+411938.8 & F1,F4,F5       & BB & $1.00^{(0.08)}_{(0.08)}$ & $65.2^{(6.8)}_{(6.2)}$        & $15.1^{(0.62)}_{(0.58)}$ & $0.00$   & N/A      & \checkmark \\
  &             &            & PL & $0.74^{(0.08)}_{(0.08)}$ & $6.75^{(1.47)}_{(1.21)}$      & $2.08^{(0.72)}_{(0.72)}$ & $0.00$   & N/A      & \checkmark \\ \tableline
24&004336.63+410811.8 & F1,F4,F5       & BB & $0.70^{(0.09)}_{(0.09)}$ & $64.8^{(7.7)}_{(6.9)}$        & $12.6^{(0.52)}_{(0.48)}$ & $0.00$   & N/A      & \checkmark \\
     &          &            & PL & $0.39^{(0.09)}_{(0.09)}$ & $2.17^{(0.54)}_{(0.44)}$      & $3.05^{(0.85)}_{(0.84)}$ & $0.00$   & N/A      & \checkmark \\ \tableline
25(fg)&004240.73+411004.3 & F1,F4,F5       & BB & $0.00^{(0.02)}_{(0.05)}$ & $1.01^{(0.04)}_{(0.04)}$     & $7.6^{(0.08)}_{(0.08)}$  & $67.06$  & N/A      &                       \\
 &              &            & PL & $0.00^{(0.02)}_{(0.06)}$ & $6.58^{(0.55)}_{(0.51)}$      & $5.69^{(0.32)}_{(0.32)}$ & $179.77$ & N/A      &                       \\ \tableline
26&004218.63+411401.7 & F1,F4,F5       & BB & $0.47^{(0.04)}_{(0.04)}$ & $77.2^{(4.2)}_{(4.0)}$        & $14.2^{(0.29)}_{(0.28)}$ & $0.00$   & N/A      & \checkmark \\
   &            &            & PL & $0.18^{(0.04)}_{(0.04)}$ & $6.57^{(0.74)}_{(0.66)}$      & $2.40^{(0.41)}_{(0.41)}$ & $0.00$   & N/A      & \checkmark \\ \tableline
27&004314.36+410721.1 & F1,F4,F5       & BB & $0.35^{(0.05)}_{(0.05)}$ & $59.9^{(3.6)}_{(3.4)}$        & $14.1^{(0.31)}_{(0.30)}$ & $0.00$   & N/A      & \checkmark \\
  &             &            & PL & $0.06^{(0.04)}_{(0.04)}$ & $3.79^{(0.46)}_{(0.41)}$      & $2.44^{(0.45)}_{(0.44)}$ & $0.00$   & N/A      & \checkmark \\ \tableline
28&004226.14+412551.9 & F1,F4,F5       & BB & $0.31^{(0.12)}_{(0.12)}$ & $9.7^{(1.7)}_{(1.5)}$         & $19.1^{(1.48)}_{(1.29)}$ & $0.00$   & N/A      & \checkmark \\
        &       &            & PL & $0.10^{(0.12)}_{(0.12)}$ & $0.48^{(0.18)}_{(0.13)}$      & $1.13^{(1.24)}_{(1.21)}$ & $0.00$   & N/A      & \checkmark \\ \tableline
30&004244.35+411608.6 & F1,F4,F5       & BB & $0.39^{(0.04)}_{(0.04)}$ & $58.8^{(3.1)}_{(2.9)}$        & $16.6^{(0.36)}_{(0.35)}$ & $0.00$   & N/A      & \checkmark \\
   &            &            & PL & $0.15^{(0.04)}_{(0.04)}$ & $8.84^{(0.94)}_{(0.85)}$      & $1.69^{(0.39)}_{(0.39)}$ & $0.00$   & N/A      & \checkmark \\ \tableline
31&004315.41+411124.6 & F1,F4,F5       & BB & $0.81^{(0.07)}_{(0.07)}$ & $47.5^{(4.7)}_{(4.3)}$        & $15.9^{(0.63)}_{(0.59)}$ & $0.00$   & N/A      & \checkmark \\
       &        &            & PL & $0.56^{(0.07)}_{(0.07)}$ & $4.73^{(0.97)}_{(0.81)}$      & $1.85^{(0.70)}_{(0.69)}$ & $0.00$   & N/A      & \checkmark \\ \tableline
32&004438.33+412530.3 & F1,F3,F4,F5    & BB & $0.09^{(0.07)}_{(0.07)}$ & $10.2^{(1.0)}_{(0.9)}$        & $21.0^{(0.96)}_{(0.89)}$ & $0.30$   & $0.30$   & \checkmark \\
      &         &            & PL & $0.00^{(0.07)}_{(0.07)}$ & $1.04^{(0.21)}_{(0.18)}$      & $0.77^{(0.72)}_{(0.71)}$ & $0.97$   & $0.97$   &         \checkmark               \\ \tableline
33&004412.20+413148.2 & F1,F4          & BB & $0.30$                   & $9.3^{(0.8)}_{(0.8)}$         & $26.3^{(1.24)}_{(1.14)}$ & $0.00$   & N/A      & \checkmark \\
       &        &            & PL & $0.30$                   & $2.40^{(0.42)}_{(0.36)}$      & $0.23^{(0.62)}_{(0.61)}$ & $0.00$   & N/A      & \checkmark \\ \tableline
34&004451.08+412904.9 & F1,F3,F4,F5    & BB & $0.00^{(0.03)}_{(0.03)}$ & $26.3^{(1.1)}_{(1.0)}$        & $21.0^{(0.40)}_{(0.39)}$ & $4.46$   & $4.46$   &                       \\
      &         &            & PL & $0.00^{(0.03)}_{(0.03)}$ & $8.89^{(0.70)}_{(0.65)}$      & $0.72^{(0.31)}_{(0.31)}$ & $15.04$  & $15.04$  &                       \\ \tableline
35(fg) &004118.65+405158.6 & F1,F3,F4,F5    & BB & $0.00^{(0.04)}_{(0.03)}$ & $0.716^{(0.025)}_{(0.025)}$    & $6.0^{(0.05)}_{(0.05)}$  & $6.15$   & $6.15$   &                       \\
       &        &            & PL & $0.00^{(0.03)}_{(0.03)}$ & $5.14^{(0.38)}_{(0.35)}$      & $7.84^{(0.26)}_{(0.26)}$ & $12.10$  & $12.10$  &                       \\ \tableline
36&004206.12+410247.7 & F1,F3,F4,F5    & BB & $0.21^{(0.09)}_{(0.09)}$ & $8.7^{(1.1)}_{(1.0)}$         & $21.2^{(1.33)}_{(1.19)}$ & $1.92$   & $1.92$   &                       \\
      &         &            & PL & $0.03^{(0.09)}_{(0.09)}$ & $0.63^{(0.17)}_{(0.14)}$      & $0.77^{(0.94)}_{(0.93)}$ & $0.89$   & $0.89$   & \checkmark \\ \tableline
37\tablenotemark{$d$} &004210.36+405149.2 & F1,F3,F4,F5    & BB & $1.02^{(0.06)}_{(0.06)}$ & $40.7^{(3.4)}_{(3.1)}$        & $19.3^{(0.79)}_{(0.73)}$ & $1.25$   & $1.25$   & \checkmark \\
      &         &            & PL & $0.87^{(0.06)}_{(0.06)}$ & $10.73^{(1.83)}_{(1.56)}$     & $0.91^{(0.64)}_{(0.64)}$ & $3.48$   & $3.48$   &                       \\ \tableline
38(fg) &004143.44+410504.7 & F1,F3,F4,F5    & BB & $0.00^{(0.02)}_{(0.02)}$ & $1.92^{(0.05)}_{(0.05)}$    & $6.0^{(0.03)}_{(0.03)}$  & $10.00$  & $10.00$  &                       \\
     &          &            & PL & $0.00^{(0.01)}_{(0.04)}$ & $22.65^{(1.24)}_{(1.18)}$     & $8.13^{(0.22)}_{(0.22)}$ & $110.64$ & $110.64$ &                       \\ \tableline
39&004126.25+405326.0 & F1,F3,F4,F5    & BB & $0.00^{(0.06)}_{(0.06)}$ & $28.7^{(2.3)}_{(2.2)}$        & $13.2^{(0.38)}_{(0.36)}$ & $3.18$   & $3.18$   &                       \\
        &       &            & PL & $0.00^{(0.05)}_{(0.07)}$ & $1.30^{(0.22)}_{(0.19)}$      & $2.46^{(0.68)}_{(0.68)}$ & $8.49$   & $8.49$   &                       \\ \tableline
40&004200.45+405942.7 & F1,F3,F4,F5    & BB & $0.06^{(0.10)}_{(0.10)}$ & $8.2^{(1.1)}_{(1.0)}$         & $18.8^{(1.17)}_{(1.05)}$ & $1.64$   & $1.64$   &     \checkmark                   \\
    &           &            & PL & $0.00^{(0.09)}_{(0.10)}$ & $0.48^{(0.14)}_{(0.11)}$      & $1.09^{(1.06)}_{(1.04)}$ & $0.86$   & $0.86$   & \checkmark \\ \tableline
41&004205.62+405713.4 & F1,F3,F4,F5    & BB & $0.45^{(0.03)}_{(0.03)}$ & $54.8^{(1.9)}_{(1.8)}$        & $19.7^{(0.34)}_{(0.33)}$ & $0.00$   & $0.00$   & \checkmark \\
  &             &            & PL & $0.29^{(0.03)}_{(0.03)}$ & $20.17^{(1.40)}_{(1.31)}$     & $0.92^{(0.28)}_{(0.28)}$ & $3.55$   & $3.55$   &                       \\ \tableline
42&004234.98+404838.8 & F1,F3,F4,F5    & BB & $0.00^{(0.04)}_{(0.05)}$ & $60.3^{(3.9)}_{(3.6)}$        & $11.3^{(0.23)}_{(0.22)}$ & $5.26$   & $5.26$   &                       \\
    &           &            & PL & $0.00^{(0.04)}_{(0.06)}$ & $2.26^{(0.30)}_{(0.27)}$      & $3.17^{(0.54)}_{(0.54)}$ & $22.55$  & $22.55$  &                       \\ \tableline
43&004241.80+405154.6 & F1,F3,F4,F5    & BB & $0.00^{(0.04)}_{(0.03)}$ & $465.7^{(18.9)}_{(18.2)}$     & $7.8^{(0.08)}_{(0.08)}$  & $5.00$   & $5.00$   &                       \\
    &           &            & PL & $0.00^{(0.03)}_{(0.04)}$ & $7.17^{(0.60)}_{(0.55)}$      & $5.63^{(0.31)}_{(0.31)}$ & $13.77$  & $13.77$  &                       \\ \tableline
44&003823.84+401250.0 & F1,F2          & BB & $0.30$                   & $12.8^{(1.5)}_{(1.3)}$        & $18.7^{(0.83)}_{(0.77)}$ & $0.00$   & N/A      & \checkmark \\
 &              &            & PL & $0.30$                   & $1.76^{(0.42)}_{(0.34)}$      & $2.20^{(0.88)}_{(0.86)}$ & $0.00$   & N/A      & \checkmark \\ \tableline
45(fg) &004432.46+410533.7 & F1,F3,F4,F5    & BB & $1.66^{(0.03)}_{(0.03)}$ & $6.34^{(0.24)}_{(0.23)}$ & $6.1^{(0.05)}_{(0.05)}$  & $3.25$   & $3.25$   &                       \\
   &            &            & PL & $1.19^{(0.03)}_{(0.03)}$ & $76.59^{(5.82)}_{(5.41)}$     & $8.61^{(0.25)}_{(0.25)}$ & $7.63$   & $7.63$   &                       \\ \tableline
46&004410.16+411830.4 & F1,F3,F4,F5    & BB & $0.79^{(0.10)}_{(0.10)}$ & $17.9^{(2.4)}_{(2.1)}$        & $21.0^{(1.43)}_{(1.27)}$ & $2.04$   & $2.04$   & \checkmark \\
 &              &            & PL & $0.70^{(0.10)}_{(0.10)}$ & $3.29^{(0.93)}_{(0.73)}$      & $0.59^{(1.10)}_{(1.09)}$ & $3.72$   & $3.72$   &                       \\ \tableline
47&004356.41+412202.2 & F1,F4,F5       & BB & $0.35^{(0.09)}_{(0.09)}$ & $102.4^{(11.9)}_{(10.7)}$     & $8.7^{(0.29)}_{(0.27)}$  & $0.00$   & N/A      & \checkmark \\
 &              &            & PL & $0.00^{(0.09)}_{(0.09)}$ & $0.38^{(0.10)}_{(0.08)}$      & $5.39^{(0.84)}_{(0.84)}$ & $0.10$   & N/A      &    \checkmark                   \\ \tableline
48&004229.72+405247.8 & F1,F4,F5       & BB & $1.52^{(0.06)}_{(0.06)}$ & $387.3^{(27.3)}_{(25.5)}$     & $9.6^{(0.23)}_{(0.22)}$  & $0.00$   & N/A      & \checkmark \\
&               &            & PL & $1.14^{(0.06)}_{(0.06)}$ & $10.37^{(1.51)}_{(1.32)}$     & $4.76^{(0.44)}_{(0.44)}$ & $0.00$   & N/A      & \checkmark \\ \tableline
49&004257.63+412137.6 & F1,F4,F5       & BB & $0.40^{(0.10)}_{(0.10)}$ & $265.9^{(30.2)}_{(27.1)}$     & $7.4^{(0.21)}_{(0.20)}$  & $0.00$   & N/A      & \checkmark \\
&               &            & PL & $0.00^{(0.09)}_{(0.10)}$ & $0.48^{(0.12)}_{(0.10)}$      & $6.85^{(0.76)}_{(0.76)}$ & $0.11$   & N/A      &    \checkmark                   \\ \tableline
50&004337.27+411443.6 & F1,F4,F5       & BB & $0.41^{(0.11)}_{(0.11)}$ & $102.6^{(14.5)}_{(12.7)}$     & $9.0^{(0.36)}_{(0.33)}$  & $0.00$   & N/A      & \checkmark \\
 &              &            & PL & $0.00^{(0.11)}_{(0.11)}$ & $0.43^{(0.13)}_{(0.10)}$      & $5.27^{(0.99)}_{(0.99)}$ & $0.00$   & N/A      & \checkmark \\ \tableline
51(fg) &004301.62+411052.7 & F1,F4,F5       & BB & $0.47^{(0.05)}_{(0.05)}$ & $1.49^{(0.08)}_{(0.08)}$    & $5.9^{(0.07)}_{(0.07)}$  & $0.00$   & N/A      & \checkmark \\
    &           &            & PL & $0.00^{(0.05)}_{(0.05)}$ & $1.32^{(0.16)}_{(0.14)}$      & $9.03^{(0.38)}_{(0.38)}$ & $0.34$   & N/A      &   \checkmark                    \\ \tableline
53(fg) &004409.84+412813.9 & F4,F5          & BB & $0.30$                   & $1.98^{(0.18)}_{(0.16)}$  & $5.4^{(0.09)}_{(0.09)}$  & $0.00$   & N/A      & \checkmark \\
  &             &            & PL & $0.30$                   & $1.20^{(0.23)}_{(0.19)}$      & $7.84^{(0.61)}_{(0.61)}$ & $0.00$   & N/A      & \checkmark \\ \tableline
54&004227.41+405936.1 & F4,F5          & BB & $0.30$                   & $178.1^{(22.1)}_{(19.7)}$     & $7.2^{(0.23)}_{(0.22)}$  & $0.00$   & N/A      & \checkmark \\
   &            &            & PL & $0.30$                   & $0.71^{(0.19)}_{(0.15)}$      & $5.07^{(0.82)}_{(0.82)}$ & $0.00$   & N/A      & \checkmark \\ \tableline
55&004137.85+410108.2 & F4,F5          & BB & $0.30$                   & $19.2^{(1.8)}_{(1.6)}$        & $19.1^{(1.04)}_{(0.95)}$ & $0.00$   & N/A      & \checkmark \\
  &             &            & PL & $0.30$                   & $3.82^{(0.72)}_{(0.60)}$      & $0.21^{(0.77)}_{(0.77)}$ & $0.00$   & N/A      & \checkmark \\ \tableline
56&004115.38+410102.5 & F4,F5          & BB & $0.30$                   & $65.0^{(6.9)}_{(6.2)}$        & $10.2^{(0.38)}_{(0.35)}$ & $0.00$   & N/A      & \checkmark \\
&               &            & PL & $0.30$                   & $1.71^{(0.38)}_{(0.32)}$      & $2.72^{(0.78)}_{(0.78)}$ & $0.00$   & N/A      & \checkmark \\ \tableline
57&004301.42+413017.3 & F4,F5          & BB & $0.30$                   & $247.0^{(27.3)}_{(24.6)}$     & $6.9^{(0.19)}_{(0.18)}$  & $0.00$   & N/A      & \checkmark \\
&               &            & PL & $0.30$                   & $0.84^{(0.20)}_{(0.16)}$      & $5.46^{(0.73)}_{(0.73)}$ & $0.00$   & N/A      & \checkmark \\ \tableline
60&004318.45+411142.2 & F1,F4,F5       & BB & $1.13^{(0.11)}_{(0.11)}$ & $28.7^{(4.2)}_{(3.7)}$        & $19.9^{(1.39)}_{(1.23)}$ & $0.00$   & N/A      & \checkmark \\
 &              &            & PL & $0.93^{(0.11)}_{(0.11)}$ & $5.18^{(1.61)}_{(1.23)}$      & $0.93^{(0.99)}_{(0.98)}$ & $0.00$   & N/A      & \checkmark \\  \tableline
\enddata
\tablenotetext{a}{For sources detected in only two UVIT filters, the $A_V$ value is fixed at $0.30$}
\tablenotetext{b}{For sources identified as foreground stars with distances (GD in Table 1), the radius is calculated using the best-fit Gaia distance and the source number has (fg) appended.}
\tablenotetext{c}{The reduced chi-square $\chi^2_{\nu}$ is given by $\chi^2$ divided by the degrees of freedom, where the degrees of freedom is taken as $\#$ of UVIT filters $-$ 3.}
\tablenotetext{d}{If source 37 is a foreground star and at the Gaia DR2 distance, the radius is 0.04 $R_{\odot}$.}
\tablecomments{Only sources detected in two or more UVIT filters (minimum requirement for fitting) are included in this table.}
\end{deluxetable*}

\clearpage
\end{document}